\documentclass{article}%
\usepackage{amsfonts}
\usepackage{amsmath}
\usepackage{fancybox}
\usepackage{appendix}%
\usepackage{amssymb}%
\usepackage[dvips]{graphicx}
\usepackage{graphicx}
\usepackage{psfrag}
\providecommand{\U}[1]{\protect\rule{.1in}{.1in}}
\newtheorem{theorem}{Theorem}
\newtheorem{acknowledgement}[theorem]{Acknowledgement}

\begin{document}

\title{Gravitational lensing and frame dragging of light in the Kerr-Newman and the Kerr-Newman (anti) de
Sitter black hole spacetimes}
\author{G. V. Kraniotis \footnote{email: gkraniot@cc.uoi.gr}\\
University of Ioannina, Physics Department \\ Section of
Theoretical Physics, GR- 451 10, Greece \\
}

 \maketitle

\begin{abstract}
The null geodesics that describe photon orbits in the spacetime of
a rotating electrically charged black hole (Kerr-Newman) are
solved exactly including the contribution from the cosmological
constant. We derive elegant closed form solutions for relativistic
observables such as the deflection angle and frame dragging effect
that a light ray experiences in the gravitational fields (i) of a
Kerr-Newman black hole and (ii) of a Kerr-Newman-de Sitter black
hole. We then solve the more involved problem of treating a
Kerr-Newman black hole as a gravitational lens, i.e. a KN black
hole along with a static source of light and a static observer
both located far away but otherwise at arbitrary positions in
space. For this model, we derive the analytic solutions of the
lens equations in terms of Appell and Lauricella hypergeometric
functions and the Weierstra\ss \ modular form. The exact solutions
derived for null, spherical polar and non-polar orbits, \ are
applied for the calculation of frame dragging for the orbit of a
photon around the galactic centre, assuming that the latter is a
Kerr-Newman black hole. We also apply the exact solution for the
deflection angle of an equatorial light ray in the gravitational
field of a Kerr-Newman black hole for the calculation of bending
of light from the gravitational field of the galactic centre for
various values of the Kerr parameter, electric charge and impact
factor. In addition, we derive expressions for the Maxwell tensor
components for a Zero-Angular-Momentum-Observer (ZAMO) in the
Kerr-Newman-de Sitter spacetime.
\end{abstract}

\section{Introduction}

One of the most fundamental exact non-vacuum solutions of the
gravitational field equations of general relativity is the
Kerr-Newman black hole \cite{Newman}. The Kerr-Newman (KN) exact
solution describes the curved spacetime geometry surrounding a
charged, rotating black hole and it solves the coupled system of
differential equations for the gravitational and electromagnetic
fields \cite{Newman} (see also \cite{HansOhanian}).

 The KN exact solution generalized the Kerr
solution \cite{KerrR}, which describes the curved spacetime
geometry around a rotating black hole, to include a net electric
charge carried by the black hole.

A more realistic description should include the cosmological
constant \cite{Supern}, \cite{GVKSWB},\cite{GKSWMERCURY}
\cite{KraniotisGVlightII},
\cite{Claus},\cite{Zdenek},\cite{Sultana}.
 Taking into account the contribution from the cosmological
constant $\Lambda,$ the generalization of the Kerr-Newman solution
is described by the Kerr-Newman de Sitter $($KNdS$)$ metric
element which in Boyer-Lindquist (BL) coordinates is given by
\cite{Stuchlik1},\cite{Carter},\cite{GrifPod}:

\begin{align}
\mathrm{d}s^{2}  & =\frac{\Delta_{r}^{KN}}{\Xi^{2}\rho^{2}}(c\mathrm{d}%
t-a\sin^{2}\theta\mathrm{d}\phi)^{2}-\frac{\rho^{2}}{\Delta_{r}^{KN}%
}\mathrm{d}r^{2}-\frac{\rho^{2}}{\Delta_{\theta}}\mathrm{d}\theta
^{2}\nonumber\\
& -\frac{\Delta_{\theta}\sin^{2}\theta}{\Xi^{2}\rho^{2}}(ac\mathrm{d}%
t-(r^{2}+a^{2})\mathrm{d}\phi)^{2}%
\end{align}%
\begin{equation}
\Delta_{\theta}:=1+\frac{a^{2}\Lambda}{3}\cos^{2}\theta,
\;\Xi:=1+\frac {a^{2}\Lambda}{3},
\end{equation}

\begin{equation}
\Delta_{r}^{KN}:=\left(  1-\frac{\Lambda}{3}r^{2}\right)  \left(  r^{2}%
+a^{2}\right)  -2\frac{GM}{c^{2}}r+\frac{Ge^{2}}{c^4},
\end{equation}

\begin{equation}
\rho^{2}=r^{2}+a^{2}\cos^{2}\theta,
\end{equation}
where $a,M,e,$ denote the Kerr parameter, mass and electric charge
of the black hole, respectively. Also $G$ denotes the
gravitational constant of Newton and $c$ the speed of light. This
is accompanied by a non-zero electromagnetic field
$F=\mathrm{d}A,$ where the vector potential (in units $G=c=1$) is
\cite{ZST},\cite{GrifPod}:
\begin{equation}
A=-\frac{er}{\Xi(r^{2}+a^{2}\cos^{2}\theta)}(\mathrm{d}t-a\sin^{2}\theta
\mathrm{d}\phi).
\end{equation}
As a consequence the $2$-form of the electromagnetic field is
computed to be:
\begin{align}
F  & =-\frac{e[-r^{2}+a^{2}\cos^{2}\theta]}{\Xi(r^{2}+a^{2}\cos^{2}\theta)^{2}%
}\mathrm{d}r\wedge\mathrm{d}t-\frac{2era^{2}\cos\theta\sin\theta}{\Xi(r^{2}%
+a^{2}\cos^{2}\theta)^{2}}\mathrm{d}\theta\wedge\mathrm{d}t\nonumber\\
& +\frac{a\sin^{2}\theta
e[-r^{2}+a^{2}\cos^{2}\theta]}{\Xi(r^{2}+a^{2}\cos
^{2}\theta)^{2}}\mathrm{d}r\wedge\mathrm{d}\phi+\frac{2era\cos\theta\sin
\theta(r^{2}+a^{2})}{\Xi(r^{2}+a^{2}\cos^{2}\theta)^{2}}\mathrm{d}\theta
\wedge\mathrm{d}\phi.\nonumber\\
&
\end{align}
In appendix C, we compute the electric and
magnetic fields for the Kerr-Newman-de Sitter spacetime as
perceived by the Zero-Angular-Momentum-Observers (ZAMO) or else
known as the locally-non-rotating frame \cite{Bardeen}.

 For the surrounding spacetime to represent a black hole, i.e.
the singularity surrounded by the horizon, the electric charge and
angular
momentum $J$ must be restricted by the relation:
\begin{equation}
\fbox{$\dfrac{GM}{c^{2}}\geq\left[  \left(  \dfrac{J}{Mc}\right)  ^{2}%
+\dfrac{Ge^{2}}{c^{4}}\right]  ^{1/2}$}\Leftrightarrow
\end{equation}%
\begin{align}
\dfrac{GM}{c^{2}}  & \geq\left[  a^{2}+\dfrac{Ge^{2}}{c^{4}}\right]
^{1/2}\Rightarrow\\
e^{2}  & \leq GM^{2}(1-a^{\prime2})\label{FundEcGMa}%
\end{align}
where in the last inequality $a^{\prime}=\frac{a}{GM/c^2}$ denotes
a dimensionless Kerr parameter.  The KN(a)dS dynamical system of
geodesics is a completely integrable system \footnote{By solving
the Hamilton-Jacobi equation by the method of separation of
variables.} as was shown in
\cite{BCAR},\cite{Stuchlik1},\cite{ZST},\cite{GVKraniotisMPI} \
and the geodesic differential equations take the form:
\begin{align}
\int\frac{\mathrm{d}r}{\sqrt{R^{\prime}}}  &
=\int\frac{\mathrm{d}\theta
}{\sqrt{\Theta^{\prime}}},\label{lensKN1}\\
\rho^{2}\frac{\mathrm{d}\phi}{\mathrm{d}\lambda}  & =-\frac{\Xi^{2}}%
{\Delta_{\theta}\sin^{2}\theta}\left(  aE\sin^{2}\theta-L\right)  +\frac
{a\Xi^{2}}{\Delta_{r}^{KN}}\left[  (r^{2}+a^{2})E-aL\right]  ,\label{azimueq}%
\\
c\rho^{2}\frac{\mathrm{d}t}{\mathrm{d}\lambda}  & =\frac{\Xi^{2}(r^{2}%
+a^{2})[(r^{2}+a^{2})E-aL]}{\Delta_{r}^{KN}}-\frac{a\Xi^{2}(aE\sin^{2}%
\theta-L)}{\Delta_{\theta}},\\
\rho^{2}\frac{\mathrm{d}r}{\mathrm{d}\lambda}  & =\pm\sqrt{R^{\prime}},\\
\rho^{2}\frac{\mathrm{d}\theta}{\mathrm{d}\lambda}  & =\pm\sqrt%
{\Theta^{\prime}},\label{polareq}%
\end{align}
where%
\begin{align}
R^{\prime}  & :=\Xi^{2}\left[  (r^{2}+a^{2})E-aL\right]  ^{2}-\Delta_{r}%
^{KN}(\mu^{2}r^{2}+Q+\Xi^{2}(L-aE)^{2}),\\
\Theta^{\prime}  & :=\left[  Q+(L-aE)^{2}\Xi^{2}-\mu^{2}a^{2}\cos^{2}%
\theta\right]  \Delta_{\theta}-\Xi^{2}\frac{\left(  aE\sin^{2}\theta-L\right)
^{2}}{\sin^{2}\theta}.
\end{align}
Null geodesics are derived by setting $\mu=0.$
In the following we use geometrized units, $G=c=1$, unless it is stipulated otherwise.

Despite the theoretical significance of Kerr-Newman and
Kerr-Newman-(anti) de Sitter black holes, and their possible
application to relativistic astrophysics, there is a scarcity of
exact analytic results in the integration of the test particle and
photon geodesic equations particularly in the latter case.

Most of the studies of photon trajectories in the KN and KN(a)dS
spacetimes have focused in the investigation of the allowed
parameter space and/or the analysis of selected orbits, see for
instance \cite{Calvani} and \cite{Stuchlik1},\cite{ZdeStu} without
actually integrating the null geodesic equations. For the case of
charged particle orbits in the KN spacetime, we refer the reader
to the works of \cite{Ruffini} and \cite{EVAh} (see also
\cite{RufII}).

It is therefore the purpose of this paper, to calculate for the
\textit{first time, }the exact solutions of the null geodesics in
KN and KN(a)dS spacetimes and produce closed form solutions for
relativistic observables such as: the deflection angle and frame
dragging effect that light rays experience in the curved spacetime
geometry of KN and KN(a)dS spacetimes. Furthermore, we treat
analytically for the first time the more involved problem in which
we consider the electrically charged, rotating (KN) black hole as
a gravitational lens, i.e. a KN black hole along with a static
source of light and a static observer both located far away but
otherwise at arbitrary positions in space.

In addition, our contribution generalises in a
\textit{non-trivial} way our previous analytic results for the
geodesy of Kerr and Kerr-(anti) de Sitter black holes (uncharged
rotating black holes)
\cite{KraniotisGVlightII},\cite{KraniotisLightI}, which constitute
a special case of the more general KN and KN(a)dS black holes and
thus a comparison of the relativistic observables can be made. As
we shall see in the main body of the paper, the electric charge of
the black hole influences the geometry of spacetime and the
corresponding relativistic observables such as the deflection
angle and the frame dragging effect of light in a significant way
which in principle is measurable. It is pleasing that the theory
produced in this work is a complete theory for the propagation of
light signals in the field of rotating charged black holes: \textit{all} of its fundamental parameters enter the analytic solutions on an equal footing.

The resulting theory should also be of interest for the galactic
centre studies given the strong experimental evidence we have from
the observation of stellar orbits (in particular from the orbits
of $S-$stars in the central arcsecond of the Milky Way) and flares
that the Sagittarius A* region harbours \ a supermassive rotating
black hole with mass of 4 million solar masses
\cite{GhezA},\cite{GenzelETAL}. It will help in measuring the
fundamental properties of the black hole such as its mass, spin
and electric charge, the cosmological parameters as well as in
measuring novel relativistic effects such as the gravitational
bending of light, and the corresponding Lense-Thirring effect at
the \textit{strong field regime} of General Relativity.

Although it is not the purpose of this paper to discuss how a net electric
charge is accumulated inside the horizon of the black hole, we briefly mention
recent attempts which address the issue of the formation of charged black
holes. Indeed, we note at this point, that the authors in \cite{Ray}, have
studied the effect of electric charge in compact stars assuming that the
charge distribution is proportional to the mass density. They found solutions
with a large positive net electric charge. From the local effect of the forces
experienced on a single charged particle, they concluded that each individual
charged particle is quickly ejected from the star. This is in turn produces
\ a huge force imbalance, \ in which the gravitational force overwhelms the
repulsive Coulomb and fluid pressure forces. In such a scenario the star
collapses to form a charged black hole before all the charge leaves the system
\cite{Ray}. \ A mechanism for generating charge asymmetry that may be linked
to the formation of a charged black hole has been suggested in \cite{Cuesta}.

For some phenomenological investigations of the electric charges of some
astronomical bodies in Reissner-Nordstr\"{o}m spacetimes \footnote{Curved
spacetime geometries surrounding non-spinning charged bodies or black
holes.}and of toroidal configurations in Reissner-Nordstr\"{o}m-(anti)-de
Sitter black hole and naked singularity spacetimes see \cite{LIORIO}%
,\cite{StuchklII},\cite{VENEZ}.

The material of this paper is organized as follows: In section \ref{sppolarkn}%
, we investigate the propagation of a light signal on the sphere
and in particular we derive the exact solution of polar spherical
null geodesics in Kerr-Newman spacetime. The solution is expressed
in terms of the Weierstra\ss \ elliptic function. The amount of
frame dragging for such a photon orbit is four times the real
period of the Weierstra\ss \ function which is expressed in terms
of Gau\ss 's hypergeometric function, see eqn.(\ref{GVKFDKNp}). We
subsequently apply our exact solutions for computing the amount of
frame dragging that a polar null spherical geodesic experiences in
the gravitational field of the SgrA* galactic black hole assuming
that the latter is a Kerr-Newman black hole, for various sets of
values for the spin and electric charge of the singularity. In
section \ref{nusplambda}, we derive the exact solution for the
Lense-Thirring effect for a spherical polar null orbit in
KN-spacetime in the presence of the cosmological constant
$\Lambda.$ Our solution is expressed in terms of Appell's
hypergeometric function of two-variables $F_{1}$, see eqn
$(\ref{pollambdafmd}).$ In section \ref{snpnullKN}, we compute in
closed analytic form the amount of Lense-Thirring precession that
a spherical non-polar orbit (i.e. with impact parameter
$\Phi\neq0)$ undergoes in Kerr-Newman spacetime, in terms of
Appell's hypergeometric function \cite{Appell} and Gau\ss 's
ordinary hypergeometric function, see our eqn.
$(\ref{totalnpsphra}).$ \ We also apply our exact formula
$(\ref{totalnpsphra})$ for the calculation of frame dragging for
spherical non-polar orbits in the gravitational field of a charged
rotating (KN) black hole for various sets of values of the
physical parameters. Subsequently, in section \ref{KNADSNPSPHER}
we derive the generalization of formula $(\ref{totalnpsphra})$ in
the case where the cosmological constant is present. The closed
form analytic solution that computes the Lense-Thirring effect for
a non-polar spherical photon orbit in KN(a)dS spacetime is
expressed elegantly in terms of Lauricella's hypergeometric
function $F_{D},$ Appell's $F_{1}$ and Gau\ss 's ordinary
hypergeometric function $F,$ see Eqn.$($\ref{fmdragnpolarlam}$).$
We also compute analytically the period in the polar
$\theta-$coordinate in terms of generalized hypergeometric
functions, eqn.$(\ref{ThetaperiofLam}).$

In section \ref{LightDefKNimpor}, we solve for the \textit{first
time }in closed analytic form the important problem of the
gravitational bending of light of an equatorial null unbound
geodesic in the spacetime of a charged rotating KN black hole. Our
exact solution for the deflection angle, involves Lauricella's
hypergeometric function of three variables, Eqns.$(\ref{finadefle}%
),(\ref{ISimerDelfKN})$.

Using a uniformization of the correspondence between the
coefficients of the quartic radial polynomial-that is involved in
the calculation of the gravitational light deflection-and its
roots we provide a closed form compact analytic solution for the
four roots in terms of Weierstra\ss \ elliptic function $\wp$ and
its derivative $\wp^{\prime}.$ Subsequently, we apply thoroughly,
our analytic solution for the deflection angle
$(\ref{ISimerDelfKN})$ for the calculation of the gravitational
bending of light of an equatorial ray in the spacetime geometry of
a KN black hole. The deflection angle of an equatorial light ray
in KNdS spacetime is calculated in section \ref{defllambda}.

We then solve the more involved problem of treating a Kerr-Newman
black hole as a gravitational lens, i.e. a KN black hole along
with a static source of light and a static observer both located
far away but otherwise at arbitrary positions in space. Again, for
this model we give the analytic solutions of the lens equations in
terms of Appell and Lauricella hypergeometric functions and the
Weierstra\ss \ modular form, see eqns.$(%
\ref{Constraint1one}%
),(\ref{fakos2}),(\ref{combinlens})$. From constraints following
from the condition that a photon escapes to infinity, an equation
is derived which describes the boundary of the shadow of the
electromagnetic rotating black hole on the observer's image plane,
see Equations (\ref{iskiosfortismenisbh}). Some examples of
solutions of the lens equations, for particular values of the
physical parameters are worked out and exhibited along with the
boundary of the shadow of the KN black hole on the observer's
image plane. We use section \ref{symperasma} for our conclusions.

In Appendix A, we present the theory in which the roots of the
particular quartic radial polynomial involved in the calculations
of the photonic trajectories in the KN spacetime, are elegantly
expressed in terms of the Weirstra$\ss$ modular form $\wp$ and its
derivative $\wp^{\prime}$. In the appendix \ref{SPEriastronKN} of
section \ref{LightDefKNimpor}, we derive the closed form analytic
solution for the relativistic periastron advance of a non-circular
equatorial timelike geodesic in the KN spacetime in terms of
Appell's hypergeometric function of two-variables $F_1$ and
Gau$\ss$'s hypergeometric function $F$,
eqn.(\ref{RelativiPeriastronPrec}). We applied our exact solution
for the calculation of periastron advance for the observed orbits
of $S$-stars in the central arcsecond of the Milky Way, assuming
that the SgrA* galactic supermassive black hole is a Kerr-Newman
black hole and that the $S$-stars can be treated as neutral test
particles.

\section{Spherical polar null geodesics in Kerr-Newman
spacetime\label{sppolarkn}}

Depending on whether or not the coordinate radius $r$ is constant
along a given geodesic, the corresponding particle orbit is
characterized as spherical or non-spherical respectively. In this
section, we will concentrate on spherical polar photon orbits with
a vanishing cosmological constant, i.e. we solve exactly for the
first time null spherical polar geodesics in the Kerr-Newman
spacetime. The exact solution of the spherical timelike and null
geodesics in the Kerr and Kerr -(anti) de Sitter
spacetimes-i.e.when the electric charge of the rotating black hole
vanishes-have been derived in references
\cite{GVKraniotisMPI},\cite{KraniotisLightI}.

Assuming a zero cosmological constant, $r=r_{f}$, where $r_{f}$ is a constant
value setting $\mu=0$ and using equations $(\ref{azimueq})-(\ref{polareq})$ we
obtain%
\begin{equation}
\frac{\mathrm{d}\phi}{\mathrm{d}\theta}=\frac{\frac{aP}{\Delta^{KN}}%
-aE+\frac{L}{\sin^{2}\theta}}{\sqrt{\Theta}}%
\end{equation}
where $\Theta$ now is given by%
\begin{equation}
\Theta=Q-\cos^{2}\theta\lbrack-a^{2}E^{2}+\frac{L^{2}}{\sin^{2}\theta}]
\end{equation}
and
\begin{equation}
\Delta^{KN}:=r^{2}+a^{2}+e^{2}-2Mr.
\end{equation}
It is convenient to introduce the parameters:%
\begin{equation}
\Phi:=L/E,\mathcal{Q}:=Q/E^{2}.
\end{equation}
Now by defining the variable $z:=\cos^{2}\theta,$the previous equation can be
written as follows,%
\begin{equation}
\mathrm{d}\phi=-\frac{1}{2}\frac{\mathrm{d}z}{\sqrt{\alpha
z^{3}-z^{2}(\alpha
+\beta)+\mathcal{Q}z}}\times\left\{  \frac{aP}{\Delta^{KN}}-a+\frac{\Phi}%
{1-z}\right\} \label{initpolar}%
\end{equation}
where%
\begin{equation}
\alpha:=-a^{2},\beta:=\mathcal{Q+}\Phi^{2}.
\label{karlhumboldtinv}
\end{equation}
It has been shown that a necessary condition for an orbit to be
\textit{polar }(meaning to intersect the symmetry axis of the Kerr
field) is the vanishing of the parameter $L,$ i.e. $L=0$
\cite{Tsoubelis}. Assuming $\Phi=0,$ in equation
(\ref{initpolar}), we can transform it into the Weierstra\ss \
form of an
elliptic curve by the following substitution%
\begin{equation}
z:=-\frac{\xi+\frac{\alpha+\beta}{12}}{-\alpha/4}.
\end{equation}
Thus we obtain the integral equation%
\begin{equation}
\int\mathrm{d}\phi=\int\frac{1}{2}\frac{\mathrm{d}\xi}{\sqrt{4\xi^{3}%
-g_{2}\xi-g_{3}}}\times\left\{  \frac{aP^{\prime}}{\Delta^{KN}}-a\right\}
\end{equation}
and this orbit integral can be \textit{inverted }by the Weierstra\ss \ modular
Jacobi form%
\begin{equation}
\xi=\wp\left(\frac{-\phi+\epsilon}{A}\right)
\end{equation}
where $A:=-\frac{1}{2}\left(  \frac{aP^{\prime}}{\Delta^{KN}}-a\right)  ,$
$P^{\prime}=(r^{2}+a^{2})$ and the Weierstra\ss \ invariants take the form%
\begin{equation}
g_{2}=\frac{1}{12}(\alpha+\beta)^{2}-\frac{\mathcal{Q}\alpha}{4},g_{3}%
=\frac{1}{216}(\alpha+\beta)^{3}-\frac{\mathcal{Q}\alpha^{2}}{48}%
-\frac{\mathcal{Q}\alpha\beta}{48}\label{CarlWeierstrassInstitut}%
\end{equation}
The parameter $\xi$ is introduced for later purposes.
In terms of the original variables, the exact solution for the
polar orbit of the photon $(\Phi=0)$ takes the form:
\begin{equation}
\wp(-\phi+\epsilon)=\frac{\alpha^{\prime\prime}}{4}\cos^{2}\theta-\frac{1}%
{12}(\alpha^{\prime\prime}+\beta^{\prime\prime})=\tilde{\xi}:=\frac{\xi}{A^2},
\label{KarlW}
\end{equation}
where $\alpha^{\prime\prime}:=\alpha^{\prime}/A^{\prime2}=\frac{\alpha}%
{a^{2}A^{\prime2}}=-\frac{1}{A^{\prime2}},\beta^{\prime\prime}:=\beta^{\prime
}/A^{\prime2}=\frac{\mathcal{Q}}{a^{2}A^{\prime2}},\mathcal{Q}^{\prime\prime
}=\mathcal{Q}^{\prime}/A^{\prime2}=\frac{\mathcal{Q}}{a^{2}A^{\prime2}}$
and $\epsilon$ denotes a constant of integration
\footnote{Explicitely  eqn(\ref{KarlW}) reads:
$\tilde{\xi}=\wp(-\phi+\wp^{-1}(\tilde{\xi}_0))$, where
$\tilde{\xi}_0$ is the initial value of $\tilde{\xi}$.}. Also
$A^{\prime}$ is given by the expression%
\begin{equation}
A^{\prime}:=\frac{\frac{e^{2}}{2a^{2}}-\frac{Mr}{a^{2}}}{\frac{r^{2}%
}{a^{2}}+\frac{e^{2}}{a^{2}}+1-\frac{2Mr}{a^{2}}}.
\end{equation}
The Weierstra\ss \ invariants are then given by%
\begin{align}
g_{2}^{\prime\prime}  & =\frac{1}{12}(\alpha^{\prime\prime}+\beta
^{\prime\prime})^{2}-\frac{\mathcal{Q}^{\prime\prime}\alpha^{\prime\prime}}%
{4}=\frac{1}{12}\frac{(-a^{2}+\mathcal{Q)}^{2}}{a^{4}A^{\prime4}}%
+\frac{\mathcal{Q}}{4a^{2}A^{\prime4}},\\
g_{3}^{\prime\prime}  & =\frac{1}{432a^{6}A^{\prime6}}[-2a^{6}-3a^{4}%
\mathcal{Q}+3a^{2}\mathcal{Q}^{2}+2\mathcal{Q}^{3}].
\end{align}

The analytic expressions for the three roots $e_i,i=1,2,3$ of the
cubic in Weierstra$\ss$ form and invariants $g_{2}^{\prime\prime},g_{3}^{\prime\prime}$, can be obtained either by applying
the algorithm of Tartaglia and Cardano or directly from
eqn.(\ref{KarlW}) using the triplet of roots of the original cubic
$(z_1,z_2,z_3)=(-\mathcal{Q}/a^2,0,1)$. Either procedure yields:
\begin{align}
e_{1}  & =\frac{(a^{2}+2\mathcal{Q)}(a^{2}+e^{2}+(-2+r)r\mathcal{)}^{2}%
}{3a^{2}(e^{2}-2r)^{2}},\\
e_{2}  & =\frac{(a^{2}-\mathcal{Q})(a^{2}+e^{2}+(-2+r)r\mathcal{)}^{2}}%
{3a^{2}(e^{2}-2r)^{2}},\\
e_{3}  & =-\frac{(2a^{2}+\mathcal{Q})(a^{2}+e^{2}+(-2+r)r\mathcal{)}^{2}%
}{3a^{2}(e^{2}-2r)^{2}},
\end{align}
The roots are organized in the ascending order of magnitude: $e_{1}>e_{2}%
>e_{3}.$ Since we are assuming \ spherical orbits, there are two conditions
from the vanishing of the polynomial $R(r)$ and its first derivative.
Implementing these two conditions, expressions for the parameter $\Phi$ and
Carter's constant $\mathcal{Q}$ are obtained:%
\begin{align}
\Phi & =\frac{a^{2}+r^{2}}{a},\mathcal{Q=-}\frac{r^{4}}{a^{2}},\\
\Phi & =\frac{a^{2}M+a^{2}r+2e^{2}r-3M%
r^{2}+r^{3}}{a\left(M-r\right)},\mathcal{Q=}\frac
{-r^{2}[4a^{2}(e^{2}-Mr)+(2e^{2}+r(-3M+r))^{2}%
]}{a^{2}\left(M-r\right)^{2}}
\nonumber \label{arxikescondipcc}\\
\end{align}

However, only the second solution is physical. These are also the conditions
for the photon to escape to infinity. Equations $(\ref{arxikescondipcc}),$ for
vanishing electric charge, reduce correctly to the ones obtained in the case
of the Kerr black hole\cite{KraniotisLightI}.

The two half-periods $\omega$ and $\omega^{\prime}$ of the
Weierstra\ss \ function are given by the following Abelian
integrals \cite{Whittaker} (for discriminant $\Delta^{c}>0$)
\footnote{We also mention the following integral identity in the
definition of the real half-period $\omega$:
$\omega=\int_{e_3}^{e_2}\frac{{\rm
d}x}{\sqrt{4x^3-g_2^{\prime\prime}x-g_3^{\prime\prime}}}=\int_{e_1}^{+\infty}\frac{{\rm
d}x}{\sqrt{4x^3-g_2^{\prime\prime}x-g_3^{\prime\prime}}}$, when
all the branch points are real and the lattice rectangular
\cite{duval}. A similar integral identity holds in the definition
of $\omega^{\prime}$.}:
\begin{equation}
\omega=\int_{e_{1}}^{+\infty}\frac{\mathrm{d}t}{\sqrt{4t^{3}-g_{2}^{\prime\prime}t-g_{3}^{\prime\prime}}%
},\text{ }\omega^{\prime}=i\int_{-\infty}^{e_{3}}\frac{\mathrm{d}t}%
{\sqrt{-4t^{3}+g_{2}^{\prime\prime}t+g_{3}^{\prime\prime}}}.
\end{equation}
A closed form expression for the real half-period $\omega$ of the Jacobi
modular form of Weierstra\ss \ $\wp$ is: $\omega=\frac{1}{\sqrt{e_{1}%
-e_{3}}}\frac{\pi}{2}F\left(  \frac{1}{2},\frac{1}{2},1,\frac{e_{2}-e_{3}%
}{e_{1}-e_{3}}\right)  ,$ where $F(\alpha,\beta,\gamma,x)$ is the
hypergeometric function of Gau\ss . Thus substituting the formulae for the
roots in terms of the parameters and initial conditions, we obtain the
analytic exact result for the half-period $\omega:$
\begin{equation}
\fbox{$ \displaystyle
\omega=\frac{1}{\sqrt{\frac{(a^{2}+\mathcal{Q)(}a^{2}+e^{2}+(-2+r)r)^{2}%
}{a^{2}(e^{2}-2r)^{2}}}}\frac{\pi}{2}F\left(  \frac{1}{2},\frac{1}{2}%
,1,\frac{a^{2}}{a^{2}+\mathcal{Q}}\right) \label{GVKFDKNp}%
$}
\end{equation}
After a complete oscillation in latitude, the angle of longitude, which
determines the amount of dragging for the spherical photon polar orbit in the
general theory of relativity (GTR) for the KN black hole, increases by:%
\begin{equation}
\Delta\phi_{pKN}^{GTR}=4\omega.
\end{equation}
Assuming that the centre of the Milky Way is a rotating black hole
with a net electric charge, i.e. the structure of the spacetime
near the region SgrA*, is described by the Kerr-Newman geometry we
determined the precise frame dragging (Lense-Thirring effect) of a
null orbit with a spherical polar geometry. For the values of the
spin of the galactic centre black hole we used values inferred by
observations \cite{Genzel},\cite{Aschenbach}. Our results are displayed in Table \ref{Nuuk}.

\begin{table}[tbp] \centering
\begin{tabular}
[c]{|l|l|l|}\hline
\textbf{Parameters} & \textbf{Half-Period} & \textbf{Predicted dragging}%
\\\hline
$1)$ $a=0.52,e=0.85,$ & $\omega=0.529682$ & $\Delta\phi_{pKN}^{GTR}=2.11873$\\
$r=2.13889,\mathcal{Q}=17.9288$ &  & $=437019\operatorname{arcsec}=121.394%
\operatorname{{{}^\circ}}%
$\\
$2)$ $a=0.9939,e=0.11$ & $\omega=0.791897$ & $\Delta\phi_{pKN}^{GTR}%
=3.16759$\\
$r=2.40998,\mathcal{Q}=22.2435$ &  & $=653362\operatorname{arcsec}=181.489%
\operatorname{{{}^\circ}}%
$\\\hline
&  & \\\hline
\end{tabular}
\caption{Predictions for frame dragging from the galactic black hole for a
photonic spherical polar Kerr-Newman orbit, for the set of values for the
Kerr parameter and electric charge: $a_{\rm Gal}=0.52\frac{GM_{\rm
BH}}{c^2},e_{\rm Gal}=0.85\sqrt{G}M_{\rm BH}$ and $a_{\rm Gal}=0.9939\frac{GM_{\rm
BH}}{c^2},e_{\rm Gal}= 0.11\sqrt{G}M_{\rm BH}$.}\label{Nuuk}
\end{table}%

\subsection{Null spherical polar geodesics in Kerr-Newman black hole with the
cosmological constant and Lense-Thirring effect\label{nusplambda}}

We now derive the closed form solution for the amount of frame dragging for a
photonic spherical polar orbit in Kerr-Newman spacetime in the presence of the
cosmological constant $\Lambda,$ thus generalizing the results of the previous section.

The relevant differential equation is:%
\begin{equation}
\frac{\mathrm{d}\phi}{\mathrm{d}\theta}=\frac{\frac{-\Xi^{2}(a\sin^{2}%
\theta-\Phi)}{\Delta_{\theta}\sin^{2}\theta}+\frac{a\Xi^{2}}{\Delta_{r}^{KN}%
}[(r^{2}+a^{2})-a\Phi]}{\sqrt{[\mathcal{Q+(}\Phi-a)^{2}\Xi^{2}]\Delta_{\theta
}-\Xi^{2}a^{2}\sin^{2}\theta-\Xi^{2}\frac{\Phi^{2}}{\sin^{2}\theta}+2a\Phi
\Xi^{2}}}\label{musicofspheres}%
\end{equation}
For $\Phi=0$, and using the change of variables, $z=\cos^{2}\theta,-\frac
{1}{2}\frac{\mathrm{d}z}{\sqrt{z}}\frac{1}{\sqrt{1-z}}=\mathrm{sign(}\frac
{\pi}{2}-\theta)\mathrm{d}\theta,$ after a complete oscillation in latitude,
the angle of longitude $\Delta\phi_{pKN\Lambda}^{GTR},$ which determines the
amount of dragging for the spherical polar orbit, is given by%
\begin{align}
\Delta\phi_{pKN\Lambda}^{GTR}  & =4\int_{0}^{1}-\frac{1}{2}\frac{\mathrm{d}%
z}{\sqrt{z}\sqrt{1-z}}%
\Biggl\{%
-\frac{\Xi^{2}a}{(1+\frac{a^{2}\Lambda}{3}z)\sqrt{\mathcal{Q}+z[\mathcal{Q}%
a^{2}\frac{\Lambda}{3}+a^{2}\Xi^{3}]}}\nonumber\\
& +\frac{a\Xi^{2}(r^{2}+a^{2})}{\Delta_{r}^{KN}\sqrt{\mathcal{Q}%
+z[\mathcal{Q}a^{2}\frac{\Lambda}{3}+a^{2}\Xi^{3}]}}%
\Biggr\}%
\nonumber\\
& =-\frac{\Xi^{2}a}{2}\frac{1}{\sqrt{\mathcal{Q}}}F_{1}\left(  \frac{1}%
{2},1,\frac{1}{2},1,-\frac{a^{2}\Lambda}{3},-\frac{\mathcal{Q}a^{2}%
\frac{\Lambda}{3}+a^{2}\Xi^{3}}{\mathcal{Q}}\right)  \frac{\Gamma^{2}%
(1/2)}{\Gamma(1)}+\nonumber\\
&
\frac{a\Xi^{2}(r^{2}+a^{2})}{\Delta_{r}^{KN}\sqrt{\mathcal{Q}}}\frac
{1}{2}F\left(
\frac{1}{2},\frac{1}{2},1,-\frac{\mathcal{Q}a^{2}\frac{\Lambda
}{3}+a^{2}\Xi^{3}}{\mathcal{Q}}\right)
\frac{\Gamma^{2}(1/2)}{\Gamma
(1)}.\label{pollambdafmd}%
\end{align}
For $\Lambda=0,$ $\Delta\phi_{pKN\Lambda}^{GTR}$ reduces correctly to
$\Delta\phi_{pKN}^{GTR}.$ Indeed,%
\begin{align}
\Delta\phi_{pKN\Lambda}^{GTR}\underset{\Lambda=0}{\rightarrow}\Delta\phi
_{pKN}^{GTR}  & =\frac{\pi}{2\sqrt{\mathcal{Q}}}F\left(  \frac{1}{2}%
,\frac{1}{2},1,-\frac{a^{2}}{\mathcal{Q}}\right)  \left\{  \frac
{-a(e^{2}-2Mr)}{r^{2}+a^{2}+e^{2}-2Mr}\right\}
.\nonumber\\
& \label{fdpkn}%
\end{align}
For zero electric charge, equation $(\ref{fdpkn}),$ correctly reduces to the
analytic solution for frame dragging that a polar null spherical orbit
experiences in Kerr spacetime \cite{KraniotisLightI}.

\section{Spherical non-polar null geodesics in Kerr-Newman spacetime
\label{snpnullKN}}

In this section and assuming vanishing cosmological constant, we
shall derive for the first time the solution in closed analytic
form for the amount of frame dragging that a spherical non-polar
photonic orbit experiences in the gravitational field of the
KN-black hole, thereby generalising the results of
\cite{KraniotisLightI}.

For this purpose we integrate the differential equation for the azimuth
$(\ref{musicofspheres})$ (for $\Lambda=0$) for $\theta$ from $\pi/2$ to a
turning point of the polar polynomial. The roots $z_{m},z_{3}$ (of
$ \Theta(\theta)=0$%
) are expressed in terms of the integrals of motion and the cosmological
constant by the expressions:
\begin{equation}
z_{3,m}=\frac{\mathcal{Q}+\Phi^{2}\Xi^{2}-H^{2}\pm\sqrt{(\mathcal{Q}+\Phi
^{2}\Xi^{2}-H^{2})^{2}+4H^{2}\mathcal{Q}}}{-2H^{2}}\label{cubicroots}%
\end{equation}
and
\begin{equation}
H^{2}:=\frac{a^{2}\Lambda}{3}[\mathcal{Q}+(\Phi-a)^{2}\Xi^{2}]+a^{2}\Xi^{2}%
\end{equation}
For $\Lambda=0$, the turning points take the form:
\begin{equation}
\fbox{$\displaystyle z_m=\frac{a^2-{\cal Q}-\Phi^2+
\sqrt{4a^2 {\cal Q}+(-a^2+{\cal Q}+\Phi^2)^2}}{2a^2}, $}
\end{equation}
where the subscript \textquotedblleft m\textquotedblright\ \ stands for
\textquotedblleft\textrm{min/max\textquotedblright. }\ Consequently,the change
$\Delta\phi$ as $\theta$ goes through a quarter of a complete oscillation is:%
\begin{align}
\Delta\phi_{nPNK\frac{1}{4}}^{GTR}  & =-\frac{1}{2}\int_{0}^{z_{m}}%
\frac{\mathrm{d}z\Phi}{(1-z)\sqrt{z}\sqrt{\alpha z^{2}-z(\alpha
+\beta)+\mathcal{Q}}}\nonumber\\
& -\frac{1}{2}\mathcal{A}_{np}^{KN}\int_{0}^{z_{m}}\frac{\mathrm{d}z}%
{\sqrt{z}\sqrt{\alpha z^{2}-z(\alpha+\beta)+\mathcal{Q}}}\nonumber\\
&
=\frac{\Phi}{|a|}\frac{1}{\sqrt{z_{m}-z_{3}}}\frac{1}{1-z_{m}}\frac{\pi
}{2}F_{1}\left(  \frac{1}{2},1,\frac{1}{2},1,\frac{-z_{m}}{1-z_{m}}%
,\frac{z_{m}}{z_{m}-z_{3}}\right)  +\nonumber\\
&
+\frac{1}{2|a|}\mathcal{A}_{np}^{KN}\sqrt{\frac{1}{z_{m}-z_{3}}}\pi
F\left(
\frac{1}{2},\frac{1}{2},1,\frac{z_{m}}{z_{m}-z_{3}}\right)
.\label{QuartenpsherKN}%
\end{align}
where%
\begin{equation}
\mathcal{A}_{np}^{KN}=\frac{-a^{2}\Phi-a\left( e^{2}-2Mr\right)
}{\Delta^{KN}}.
\end{equation}
The total change in azimuth is:%
\begin{equation}
\Delta\phi_{nPNK}^{GTR}:=4\times\Delta\phi_{nPNK\frac{1}{4}}^{GTR}%
\label{totalnpsphra}%
\end{equation}
Orbits with $\Delta\phi_{nPNK}^{GTR}>0$ are called \textit{prograde }and
orbits with $\Delta\phi_{nPNK}^{GTR}<0$ are called \textit{retrograde.} The
differential equation relevant for the time integration is
\begin{equation}
\frac{\mathrm{d}t}{\mathrm{d}\theta}=\frac{(r^{2}+a^{2})[(r^{2}+a^{2})-\Phi
a]}{\Delta^{KN}\sqrt{\Theta}}+\frac{a\Phi}{\sqrt{\Theta}}-\frac
{a^{2}\sin^{2}\theta}{\sqrt{\Theta}}%
\end{equation}
Integrating for $\theta$ from $\pi/2$ to a turning point we obtain%
\begin{align}
t  & =\left\{  \frac{(r^{2}+a^{2})[(r^{2}+a^{2})-\Phi a]}{\Delta^{KN}}%
+a\Phi\right\}  \frac{1}{2|a|}\sqrt{\frac{1}{z_{m}-z_{3}}}\pi
F\left(
\frac{1}{2},\frac{1}{2},1,\frac{z_{m}}{z_{m}-z_{3}}\right) \nonumber\\
& -\frac{a^{2}}{|a|}\frac{(1-z_m)z_m}{\sqrt{z_m^2}}\frac{1}{\sqrt{z_m-z_3}}\frac{\pi}{2}F_{1}\left(  \frac{1}{2},-1,\frac{1}%
{2},1,\frac{-z_{m}}{1-z_{m}},\frac{z_{m}}{z_{m}-z_{3}}\right)  .
\end{align}
Assuming that the galactic centre region SgrA* harbours a supermassive
Kerr-Newman black hole we computed the frame-dragging that a null non-polar
spherical orbit experiences in a such a curved spacetime geometry. Our results
are displayed in tables \ref{NPSKNFDA052},\ref{NPKNA09939NPSnull}.%

\begin{table}[tbp] \centering
\begin{tabular}
[c]{|l|l|}\hline
\textbf{Parameters} & \textbf{Predicted Frame-dragging}\\\hline
$\Phi=1,\mathcal{Q}=13.8317,r=1.88356$ & $\Delta\phi_{nPNK}^{GTR}%
=9.05651=1.87\times10^{6}\operatorname{arcsec}=518.9%
\operatorname{{{}^\circ}}%
$\\
$\Phi=-1,\mathcal{Q}=19.6421,r=2.34727$ & $\Delta\phi_{nPNK}^{GTR}%
=-4.51044=-930345\operatorname{arcsec}=-258.429%
\operatorname{{{}^\circ}}%
$\\
$\Phi=-3,\mathcal{Q}=16.4623,r-2.68908$ & $\Delta\phi_{nPNK}^{GTR}%
=-4.88855=-1.00834\times10^{6}\operatorname{arcsec}=-280.093%
\operatorname{{{}^\circ}}%
$\\\hline
\end{tabular}
\caption{Predictions for frame-dragging from a galactic electrically charged
rotating black hole for non-polar null spherical geodesics. The Kerr
parameter is $a_{Gal}=0.52\frac{GM_{\rm BH}}{c^2}$ and the black hole's
electric charge: $e_{Gal}=0.85\sqrt{G}M_{\rm{BH}}$.}\label{NPSKNFDA052}%
\end{table}%
%

\begin{table}[tbp] \centering
\begin{tabular}
[c]{|l|l|}\hline
\textbf{Parameters} & \textbf{Predicted Frame-dragging}\\\hline
$\Phi=1,\mathcal{Q}=15.9524,r=1.99709$ & $\Delta\phi_{nPNK}^{GTR}%
=10.8305=2.23395\times10^{6}\operatorname{arcsec}=620.541%
\operatorname{{{}^\circ}}%
$\\
$\Phi=-1,\mathcal{Q}=25.7661,r=2.72683$ & $\Delta\phi_{nPNK}^{GTR}%
=-3.7207=-767450\operatorname{arcsec}=-213.181%
\operatorname{{{}^\circ}}%
$\\
$\Phi=-3,\mathcal{Q}=25.7614,r=3.22929$ & $\Delta\phi_{nPNK}^{GTR}%
=-4.32172=-89148\operatorname{arcsec}=-247.616%
\operatorname{{{}^\circ}}%
$\\\hline
\end{tabular}
\caption{Predictions from a galactic electrically charged rotating black
hole for non-polar null spherical geodesics. The Kerr parameter is
$a_{Gal}=0.9939\frac{GM_{\rm BH}}{c^2}$ and the black hole's electric
charge: $e_{Gal}=0.11\sqrt{G}M_{\rm{BH}}$}\label{NPKNA09939NPSnull}%
\end{table}%

\section{Frame-dragging for spherical non-polar null orbits in
Kerr-Newman-(anti) de Sitter spacetime \label{KNADSNPSPHER}}

In this section we shall derive the \textit{first }solution in closed analytic
form for the amount of frame-dragging that a non-polar spherical null orbit
expreriences in the KN-(a)dS spacetime. The relevant differential equation is
equation $(\ref{musicofspheres}).$ We again integrate for the polar coordinate
$\theta$ from $\pi/2$ to a turning point and use the variable $z$. Indeed, we
compute the relevant integrals in closed form in terms of the multivariable
Lauricella's fourth hypergeometric function $F_{D} $ and the
Gau\ss $^{^{\prime}}$s hypergeometric function:%
\begin{equation}
\int_{\pi/2}^{\theta\min/\max}\frac{a\Xi^{2}}{\Delta_{r}^{KN}}\frac
{[(r^{2}+a^{2})-a\Phi]}{\sqrt{\Theta^{\prime}}}\mathrm{d}\theta=\frac
{a\Xi^{2}}{\Delta_{r}^{KN}}\frac{[(r^{2}+a^{2})-a\Phi]}{|H|}\sqrt{\frac
{1}{z_{m}-z_{3}}}\frac{\pi}{2}F\left(  \frac{1}{2},\frac{1}{2},1,\frac{z_{m}%
}{z_{m}-z_{3}}\right)  .
\end{equation}
Exact integration of the other term in equation $(\ref{musicofspheres})$
yields the result:%
\begin{align}
& \int_{\pi/2}^{\theta\min/\max}-\frac{\Xi^{2}}{\Delta_{\theta}\sin^{2}\theta
}\frac{(a\sin^{2}\theta-\Phi)}{\sqrt{\Theta^{\prime}}}\mathrm{d}%
\theta\nonumber\\
& =%
\Biggl\{%
\frac{\Xi^{2}\Phi}{2|H|}\frac{z_{m}}{(1-\eta z_{m})(1-z_{m})}\frac{1}%
{\sqrt{z_{m}^{2}(z_{m}-z_{3})}}F_{D}\left(  \frac{1}{2},\mbox{\boldmath${\beta}%
_{4}^{\Lambda1}$},\frac{3}{2},\mbox{\boldmath$z_{\Lambda0}^{\alpha_{1}}$}\right)  2\nonumber\\
&
-\frac{\Xi^{2}a}{2|H|}\frac{z_{m}}{\sqrt{z_{m}^{2}(z_{m}-z_{3})}}\frac
{1}{1-\eta z_{m}}F_{D}\left(  \frac{1}{2},\mbox{\boldmath${\beta}_{3}^{4}$},\frac{3}
{2},\mbox{\boldmath$z_{\Lambda0}^{\alpha_{2}}$}\right)  2
\Biggr\}
\end{align}
where the parameters beta of the Lauricella hypergeometric function are given
by the vectors:%
\begin{equation}
\mbox{\boldmath${\beta}_4^{\Lambda1}$}  =\left(
1,1,\frac{1}{2},\frac{1}{2}\right),\;\;
\mbox{\boldmath${\beta}_3^4$}=\left( 1,\frac{1}{2},\frac{1}{2}\right)
\end{equation}
and the variables by the tuples of numbers:%
\begin{align}
\mbox{\boldmath$z_{\Lambda0}^{\alpha_{1}}$}  & =\left(  \frac{\eta(-z_{m})}{1-\eta z_{m}}%
,\frac{-z_{m}}{1-z_{m}},1,\frac{z_{m}}{z_{m}-z_{3}}\right)  ,\\
\mbox{\boldmath$z_{\Lambda0}^{\alpha_{2}}$}  & =\left(  \frac{\eta(-z_{m})}{1-\eta z_{m}%
},1,\frac{z_{m}}{z_{m}-z_{3}}\right)  .
\end{align}
Also $\eta:=-a^{2}\frac{\Lambda}{3}$. Thus we get for the amount of frame
dragging:%
\begin{align}
\Delta\phi_{nPKN\Lambda}^{GTR}  & =4
\Biggl[
\frac{a\Xi^{2}}{\Delta_{r}^{KN}}\frac{[(r^{2}+a^{2})-a\Phi]}{|H|}%
\sqrt{\frac{1}{z_{m}-z_{3}}}\frac{\pi}{2}F\left(  \frac{1}{2},\frac{1}%
{2},1,\frac{z_{m}}{z_{m}-z_{3}}\right) \nonumber\\
& +\frac{\Xi^{2}\Phi}{2|H|}\frac{z_{m}}{(1-\eta
z_{m})(1-z_{m})}\frac
{1}{\sqrt{z_{m}^{2}(z_{m}-z_{3})}}F_{D}\left(
\frac{1}{2},\mbox{\boldmath${\beta
}_{4}^{\Lambda1}$},\frac{3}{2},\mbox{\boldmath$z_{\Lambda0}^{\alpha_{1}}$}\right)  2\nonumber\\
& +-\frac{\Xi^{2}a}{2|H|}\frac{z_{m}}{\sqrt{z_{m}^{2}(z_{m}-z_{3})}}%
\frac{1}{1-\eta z_{m}}F_{D}\left(  \frac{1}{2},\mbox{\boldmath${\beta}_{3}^{4}$},\frac
{3}{2},\mbox{\boldmath$z_{\Lambda0}^{\alpha_{2}}$}\right)  2
\Biggr]
\label{fmdragnpolarlam}%
\end{align}
For $\Phi=0$, equation $(\ref{fmdragnpolarlam}),$ reduces correctly to the
result of the previous section for spherical polar null geodesics in the
presence of $\Lambda.$%
\begin{align}
& \Delta\phi_{nPKN\Lambda}^{GTR}\overset{\Phi=0}{\rightarrow}4%
\Biggl[%
\frac{a\Xi^{2}}{\Delta_{r}^{KN}}\frac{(r^{2}+a^{2})}{|H|}\sqrt{\frac{H^{2}%
}{\mathcal{Q}+H^{2}}}\frac{\pi}{2}F\left(  \frac{1}{2},\frac{1}{2}%
,1,\frac{H^{2}}{\mathcal{Q}+H^{2}}\right) \nonumber\\
& -\frac{\Xi^{2}a}{|H|}\sqrt{\frac{H^{2}}{\mathcal{Q}+H^{2}}}\frac
{1}{1+\frac{a^{2}\Lambda}{3}}F_{1}\left(  \frac{1}{2},1,\frac{1}{2}%
,1,\frac{\frac{a^{2}\Lambda}{3}}{1+\frac{a^{2}\Lambda}{3}},\frac{H^{2}%
}{\mathcal{Q}+H^{2}}\right)  \frac{\Gamma\left(  \frac{3}{2}\right)
\Gamma\left(  \frac{1}{2}\right)  }{\Gamma^{2}(1)}%
\Biggr]%
\nonumber\label{orionpofd}\\
& =\Delta\phi_{pKN\Lambda}^{GTR}%
\end{align}
In going from eqn. $(\ref{fmdragnpolarlam})$ to eqn.$(\ref{orionpofd}%
),$ we used the property of Lauricella's fourth hypergeometric function
$F_{D}:$%
\begin{align}
F_{D}(\alpha,\beta,\beta^{\prime},\beta^{\prime\prime},\gamma,x,1,z)  &
=\frac{\Gamma(\gamma)\Gamma(\gamma-\alpha-\beta^{\prime})}{\Gamma
(\gamma-\alpha)\Gamma(\gamma-\beta^{\prime})}\times\nonumber\\
& F_{1}(\alpha,\beta,\beta^{\prime\prime},\gamma-\beta^{\prime},x,z).
\end{align}
In addition, in order to establish the equality of
eqn.$(\ref{orionpofd})$ and eqn.$(\ref{pollambdafmd})$ we need the
following property \footnote{Equation (\ref{PaulApp}) is easily
proved by using the integral representation of Appell's function
$F_1$ and performing the change of variables: $u=1-v$ to the
original variable of integration $u$.} of Appell's hypergeometric
function $F_1$:
\begin{equation}
\fbox{$ \displaystyle
F_{1}(\alpha,\beta,\beta^{\prime},\gamma,x,y)=(1-x)^{-\beta}(1-y)^{-\beta
^{\prime}}F_{1}\left(\gamma-\alpha,\beta,\beta^{\prime},\gamma,\frac{x}{x-1},\frac{y}%
{y-1}\right).$} \label{PaulApp}
\end{equation}

Furthermore, eqn.(\ref{orionpofd}) reduces correctly for $\Lambda=0:$%
\begin{align}
& \Delta\phi_{nPKN\Lambda}^{GTR}\text{ }\underrightarrow{\Phi=\Lambda=0}\text{
}\frac{a}{\Delta^{KN}}\frac{r^{2}+a^{2}}{|a|}\sqrt{\frac{a^{2}}%
{\mathcal{Q}+a^{2}}}\frac{\pi}{2}F\left(  \frac{1}{2},\frac{1}{2}%
,1,\frac{a^{2}}{a^{2}+\mathcal{Q}}\right) \nonumber\\
&
-\frac{a}{|a|}\sqrt{\frac{a^{2}}{\mathcal{Q}+a^{2}}}\frac{\pi}{2}F\left(
\frac{1}{2},\frac{1}{2},1,\frac{a^{2}}{a^{2}+\mathcal{Q}}\right) \nonumber\\
& =\sqrt{\frac{a^{2}}{\mathcal{Q}+a^{2}}}\frac{\pi}{2}F\left(  \frac{1}%
{2},\frac{1}{2},1,\frac{a^{2}}{a^{2}+\mathcal{Q}}\right)  \left[  \frac
{r^{2}+a^{2}}{\Delta^{KN}}-1\right]  =\sqrt{\frac{a^{2}}{\mathcal{Q}+a^{2}%
}}\frac{\pi}{2}F\left(  \frac{1}{2},\frac{1}{2},1,\frac{a^{2}}{a^{2}%
+\mathcal{Q}}\right)  \frac{2r-e^{2}}{\Delta^{KN}}\nonumber\\
& =\Delta\phi_{pKN}^{GTR}.
\end{align}

For the time integration we shall need the differential equation:
\begin{equation}
\frac{\mathrm{d}t}{\mathrm{d}\theta}=\frac{\Xi^{2}(r^{2}+a^{2})[(r^{2}%
+a^{2})-\Phi
a]}{\pm\Delta_{r}^{KN}\sqrt{\Theta^{\prime}}}-\frac{a\Xi
^{2}(a\sin^{2}\theta-\Phi)}{\pm\Delta_{\theta}\sqrt{\Theta^{\prime}}}%
\end{equation}
Again integrating exactly for the polar angle $\theta$, from $\theta=\pi/2 $
to a turning point and using the variable $z$, yields the following results:%
\begin{align}
\int_{\pi/2}^{\theta_{\min/\max}}\frac{-a^{2}\Xi^{2}\sin^{2}\theta}{\pm
\Delta_{\theta}\sqrt{\Theta^{\prime}}}\mathrm{d}\theta &
=-a^{2}\Xi
^{2}\frac{1}{2|H|}\frac{(1-z_{m})z_{m}}{(1-\eta z_{m})\sqrt{z_{m}^{2}}%
}\frac{1}{\sqrt{z_{m}-z_{3}}}2F_{D}\left(
\frac{1}{2},\mbox{\boldmath${\beta
}$}_{\mathbf{4}}^{\mathbf{11}},\frac{3}{2},\mathbf{z}_{\mathbf{\Lambda0}%
}^{\mathbf{\alpha}_{1}}\right) \nonumber\\
& =-a^{2}\Xi^{2}\frac{1}{|H|}\frac{(1-z_{m})z_{m}}{(1-\eta z_{m}%
)\sqrt{z_{m}^{2}}}\frac{1}{\sqrt{z_{m}-z_{3}}}\frac{\Gamma\left(
\frac{3}{2}\right)  \Gamma\left(  \frac{1}{2}\right)  }{\Gamma^{2}(1)}%
\times\nonumber\\
& F_{D}\left(  \frac{1}{2},\mbox{\boldmath${\beta}$}_{\mathbf{3}}^{\mathbf{10}%
},1,\mathbf{z}_{\mathbf{\Lambda0}}^{\mbox{\boldmath${\beta}_{1}$}}\right)  ,
\end{align}
where we have defined the tuples for the beta parameters of Lauricella's
function as:
\begin{equation}
\mbox{\boldmath${\beta}$}_{\mathbf{4}}^{\mathbf{11}}=\left(  1,-1,\frac{1}{2},\frac{1}%
{2}\right)  ,\mbox{\boldmath${\beta}$}_{\mathbf{3}}^{\mathbf{10}}=\left(  1,-1,\frac
{1}{2}\right)  ,
\end{equation}
and the variable tuple:%
\begin{equation}
\mathbf{z}_{\mathbf{\Lambda0}}^{\mathbf{\beta}_{1}}=\left(  \frac{\eta
(-z_{m})}{1-\eta z_{m}},\frac{-z_{m}}{1-z_{m}},\frac{z_{m}}{z_{m}-z_{3}%
}\right)  .
\end{equation}
Likewise exact integration of the term $\int_{\pi/2}^{\theta_{\min/\max}}%
\frac{a\Xi^{2}\Phi}{\pm\Delta_{\theta}\sqrt{\Theta^{\prime}}}%
\mathrm{d}\theta$ yields the analytic result:%
\begin{equation}
\int_{\pi/2}^{\theta_{\min/\max}}\frac{a\Xi^{2}\Phi}{\pm\Delta_{\theta
}\sqrt{\Theta^{\prime}}}\mathrm{d}\theta=a\Xi^{2}\Phi\frac{1}{|H|}\frac
{1}{1-\eta
z_{m}}\frac{1}{\sqrt{z_{m}-z_{3}}}\frac{\pi}{2}F_{1}\left(
\frac{1}{2},1,\frac{1}{2},1,\frac{\eta(-z_{m})}{1-\eta z_{m}},\frac{z_{m}%
}{z_{m}-z_{3}}\right)  .
\end{equation}
Thus we obtain for the period in the $\theta$ coordinate the exact result:
\begin{align}
t  & =4\times\Biggl[\frac{\Xi^{2}(r^{2}+a^{2})[(r^{2}+a^{2})-\Phi a]}{|H| \Delta_{r}^{KN}}%
\sqrt{\frac{1}{z_{m}-z_{3}}}\frac{\pi}{2}F\left(  \frac{1}{2},\frac{1}
{2},1,\frac{z_{m}}{z_{m}-z_{3}}\right) \nonumber\\
& +a\Xi^{2}\Phi\frac{1}{|H|}\frac{1}{1-\eta z_{m}}\frac{1}{\sqrt
{z_{m}-z_{3}}}\frac{\pi}{2}F_{1}\left(  \frac{1}{2},1,\frac{1}{2},1,\frac
{\eta(-z_{m})}{1-\eta z_{m}},\frac{z_{m}}{z_{m}-z_{3}}\right) \nonumber\\
& -a^{2}\Xi^{2}\frac{1}{2|H|}\frac{(1-z_{m})z_{m}}{(1-\eta z_{m}%
)\sqrt{z_{m}^{2}}}\frac{1}{\sqrt{z_{m}-z_{3}}}2F_{D}\left(  \frac{1}
{2},\mathbf{\beta}_{\mathbf{4}}^{\mathbf{11}},\frac{3}{2},\mathbf{z}
_{\mathbf{\Lambda0}}^{\mathbf{\alpha}_{1}}\right)
\Biggr]
\label{ThetaperiofLam}
\end{align}

\section{Light deflection of an equatorial unbound Kerr-Newman orbit
\label{LightDefKNimpor}}

In this section, we are going to calculate for the first time the
exact analytic solution for the bending of light for an equatorial
unbound null geodesic in the gravitational field of an
electrically charged rotating black hole.

For equatorial geodesics the parameter $\mathcal{Q}$ vanishes and the relevant
differential equation for the exact computation is the following:%

\begin{equation}
\frac{\mathrm{d}\phi}{\mathrm{d}r}=\frac{e^{2}(\Phi-a)+\Phi r^{2}-2Mr(\Phi-a)}{(\underset{=:\Delta^{KN}}{\underbrace{r^{2}+a^{2}%
+e^{2}-2Mr}})\sqrt{R}},\label{deforbit}%
\end{equation}
where the quartic radial polynomial has the form:%
\begin{equation}
R=r^{4}+r^{2}(a^{2}-\Phi^{2})+2Mr(a^{2}+\Phi^{2}-2\Phi
a)-e^2(\Phi-a)^2.\label{quarweier}%
\end{equation}
We shall compute the following integral applying the partial fractions technique:
\begin{align}
\int\mathrm{d}\phi&=\int\frac{e^{2}(\Phi-a)+\Phi
r^{2}-2Mr%
(\Phi-a)}{\Delta^{KN}\sqrt{R}}\mathrm{d}r \nonumber\\
&=\int\frac{\Phi}{\sqrt{R}}\mathrm{d}r+\int\frac{A_{+}^{eqKN}}%
{(r-r_{+})\sqrt{R}}\mathrm{d}r+\int\frac{A_{-}^{eqKN}}{(r-r_{-})\sqrt%
{R}}\mathrm{d}r
\label{radeqdeflgravkn}
\end{align}

In order to calculate the deflection angle from the previous
radial integral (\ref{radeqdeflgravkn}) we need to integrate from the
distance of closest approach (e.g., from the maximum positive root
of the quartic) to infinity. Thus
$\Delta\phi_{eKN}^{GTR}=2\int_{\alpha}^{\infty}.$ We manipulate a
bit further
the terms in equation $(\ref{deforbit}).$ In particular:%
\begin{align}
\frac{\Phi r^{2}}{\Delta^{KN}\sqrt{R}}  & =\Phi\left[  \frac{1}{\sqrt%
{R}}-\frac{(a^{2}+e^{2}-2Mr)}{\Delta^{KN}\sqrt{R}}\right]
\Rightarrow\\
\frac{\Phi r^{2}}{\Delta^{KN}\sqrt{R}}-\frac{2Mr(\Phi
-a)}{\Delta^{KN}\sqrt{R}}  &
=\frac{\Phi}{\sqrt{R}}+\frac{2Mar-\Phi(a^{2}+e^{2})}{\Delta^{KN}\sqrt{R}}.
\label{manipulativeop}
\end{align}

We organize all roots in ascending order of magnitude as
follows,
\begin{equation}
\alpha_{\mu}>\alpha_{\nu}>\alpha_{i}>\alpha_{\rho}%
\end{equation}
where $\alpha_{\mu}=\alpha_{\mu+1},\alpha_{\nu}=\alpha_{\mu+2},\alpha_{\rho
}=\alpha_{\mu}$ and $\alpha_{i}=\alpha_{\mu-i},i=1,2,3 $ and we have that
$\alpha_{\mu-1}\geq\alpha_{\mu-2}>\alpha_{\mu-3}.$ By applying the transformation
\begin{equation}
r^{\prime}=\frac{\omega z\alpha_{\mu+2}-\alpha_{\mu+1}}{\omega z-1}%
\end{equation}
or equivalently
\begin{equation}
z=\left(  \frac{\alpha_{\mu}-\alpha_{\mu+2}}{\alpha_{\mu}-\alpha_{\mu+1}%
}\right)  \left(  \frac{r^{\prime}-\alpha_{\mu+1}}{r^{\prime}-\alpha_{\mu+2}}\right)
\end{equation}
where%
\begin{equation}
\omega:=\frac{\alpha_{\mu}-\alpha_{\mu+1}}{\alpha_{\mu}-\alpha_{\mu+2}}%
\end{equation}
and after a dimensionless variable $r^{\prime}$ through
$r=r^{\prime}M_{BH}$ has been introduced, we can bring our radial
integrals into the familiar integral representation of
Lauricella's $F_{D}$ and Appell's hypergeometric function
$F_{1\text{ \ }} $of three and two variables respectively
\footnote{See Appendix B for the integral representation that the
Appell-Lauricella hypergeometric function admits. We also have the
correspondence $\alpha_{\mu+1}=\alpha,\alpha_{\mu+2}=\beta
,\alpha_{\mu-1}=r^{\prime}_{+}=\alpha_{\mu-2},\alpha_{\mu-3}=\gamma,\alpha_{\mu}%
=\delta.$}. We denote the
roots of the quartic $(\ref{quarweier})$ by
$\alpha,\beta,\gamma,\delta:\alpha>\beta>\gamma
>\delta,$ while we define:
\begin{equation}
H^{\pm}=\sqrt{\omega}(\alpha_{\mu+1}-\alpha_{\mu+2})(\alpha_{\mu-1}^{\pm
}-\alpha_{\mu+1})\sqrt{\alpha_{\mu+1}-\alpha_{\mu}}\sqrt{\alpha_{\mu
+1}-\alpha_{\mu-3}}%
\end{equation}
The radii of the (dimensionless) event $(r^{\prime}_{+})$ and Cauchy $(r^{\prime}_{-})$ horizons are given \footnote{In the usual units: $r_{\pm}=\frac{GM_{\rm BH}}{c^2}\pm\sqrt{\left(\frac{GM_{\rm BH}}{c^2}\right)^2-\left(\frac{Ge^2}{c^4}+a^2\right)}$.} by:
\begin{equation}
r_{\pm}^{\prime}=1\pm\sqrt{1-a^{2}-e^{2}}%
\end{equation}

\begin{align}
\Delta\phi_{eKN}^{GTR}  & =2%
\Biggl[%
\frac{-2A_{+}^{eqKN}\sqrt{\omega}(\alpha_{\mu+1}-\alpha_{\mu+2})}{H^{+}}%
F_{D}\left(  \frac{1}{2},\frac{1}{2},1,\frac{1}{2},\frac{3}{2},\frac{1}%
{\omega},\kappa_{+}^{\prime2},\mu^{\prime2}\right) \nonumber\\
& +\frac{A_{+}^{eqKN}\sqrt{\omega}(\alpha_{\mu+1}-\alpha_{\mu+2})}{H^{+}}%
F_{D}\left(  \frac{3}{2},\frac{1}{2},1,\frac{1}{2},\frac{5}{2},\frac{1}%
{\omega},\kappa_{+}^{\prime2},\mu^{\prime2}\right)  \frac{\Gamma
(3/2)\Gamma(1)}{\Gamma(5/2)}\nonumber\\
& +\frac{-2A_{-}^{eqKN}\sqrt{\omega}(\alpha_{\mu+1}-\alpha_{\mu+2})}{H^{-}%
}F_{D}\left(  \frac{1}{2},\frac{1}{2},1,\frac{1}{2},\frac{3}{2},\frac
{1}{\omega},\kappa_{-}^{\prime2},\mu^{\prime2}\right) \nonumber\\
& +\frac{A_{-}^{eqKN}\sqrt{\omega}(\alpha_{\mu+1}-\alpha_{\mu+2})}{H^{-}}%
F_{D}\left(  \frac{3}{2},\frac{1}{2},1,\frac{1}{2},\frac{5}{2},\frac{1}%
{\omega},\kappa_{-}^{\prime2},\mu^{\prime2}\right)  \frac{\Gamma
(3/2)\Gamma(1)}{\Gamma(5/2)}%
\Biggr]%
\nonumber\\
& +2\Phi\frac{\Gamma(1/2)\Gamma(1)}{\Gamma(3/2)}\left(  \frac{1}%
{\sqrt{(\gamma-\alpha)(\delta-\alpha)}}\right)  F_{1}\left(  \frac{1}%
{2},\frac{1}{2},\frac{1}{2},\frac{3}{2},\frac{1}{\omega},\mu^{\prime2}\right)
\nonumber\label{arxidfle}\\
&
\end{align}
where
\begin{align}
\frac{1}{\omega}  & =\frac{\alpha_{\mu}-\alpha_{\mu+2}}{\alpha_{\mu}%
-\alpha_{\mu+1}}=\frac{\delta-\beta}{\delta-\alpha},\\
\kappa_{\pm}^{\prime2}  & =\frac{\alpha_{\mu+2}-\alpha_{\mu-1}^{\pm}}%
{\alpha_{\mu+1}-\alpha_{\mu-1}^{\pm}}=\frac{\beta-r^{\prime}_{\pm}}{\alpha-r^{\prime}_{\pm}},\\
\mu^{\prime2}  & =\frac{\alpha_{\mu+2}-\alpha_{\mu-3}}{\alpha_{\mu+1}%
-\alpha_{\mu-3}}=\frac{\beta-\gamma}{\alpha-\gamma}.
\end{align}
An equivalent expression is the following:
\begin{align}
\Delta\phi_{eKN}^{GTR}  & =2%
\Biggl[%
\frac{-2A_{+}^{eqKN}\sqrt{\omega}(\alpha_{\mu+1}-\alpha_{\mu+2})}{H^{+}}%
F_{D}\left(  \frac{1}{2},\mathbf{\beta}_{\mathbf{3}}^{\mathbf{9}},\frac{3}%
{2},\mathbf{z}_{+}^{\mathbf{r}}\right) \nonumber\\
& +\frac{A_{+}^{eqKN}\sqrt{\omega}(\alpha_{\mu+1}-\alpha_{\mu+2})}{H^{+}}%
\Biggl(%
-\frac{1}{\kappa_{+}^{\prime2}}F_{1}\left(  \frac{1}{2},\mathbf{\beta}%
_{A}^{\mathbf{ra}},\frac{3}{2},\mathbf{z}_{\mathbf{A}}^{\mathbf{r}}\right)
2\nonumber\\
& +\frac{1}{\kappa_{+}^{\prime2}}F_{D}\left(  \frac{1}{2},\mathbf{\beta
}_{\mathbf{3}}^{\mathbf{9}},\frac{3}{2},\mathbf{z}_{+}^{\mathbf{r}}\right)  2%
\Biggr)%
\nonumber\\
& +\frac{-2A_{-}^{eqKN}\sqrt{\omega}(\alpha_{\mu+1}-\alpha_{\mu+2})}{H^{-}%
}F_{D}\left(  \frac{1}{2},\mathbf{\beta}_{\mathbf{3}}^{\mathbf{9}},\frac{3}%
{2},\mathbf{z}_{-}^{\mathbf{r}}\right) \nonumber\\
& +\frac{A_{-}^{eqKN}\sqrt{\omega}(\alpha_{\mu+1}-\alpha_{\mu+2})}{H^{-}}%
\Biggl(%
-\frac{1}{\kappa_{-}^{\prime2}}F_{1}\left(  \frac{1}{2},\mathbf{\beta}%
_{A}^{\mathbf{ra}},\frac{3}{2},\mathbf{z}_{\mathbf{A}}^{\mathbf{r}}\right)
2\nonumber\\
& +\frac{1}{\kappa_{-}^{\prime2}}F_{D}\left(  \frac{1}{2},\mathbf{\beta
}_{\mathbf{3}}^{\mathbf{9}},\frac{3}{2},\mathbf{z}_{-}^{\mathbf{r}}\right)  2%
\Biggr)%
\Biggr]%
\nonumber\\
& +2\Phi\frac{\Gamma(1/2)\Gamma(1)}{\Gamma(3/2)}\left(  \frac{1}%
{\sqrt{(\gamma-\alpha)(\delta-\alpha)}}\right)  F_{1}\left(  \frac{1}%
{2},\mathbf{\beta}_{A}^{\mathbf{ra}},\frac{3}{2},\mathbf{z}_{\mathbf{A}%
}^{\mathbf{r}}\right), \label{finadefle}%
\end{align}
where
\begin{align}
A_{+}^{eqKN}  & =\frac{e^{2}(\Phi-a)}{2\sqrt{1-a^{2}-e^{2}}}+\frac
{-2a(1+\sqrt{1-a^{2}-e^{2}})+\Phi(a^{2}+e^{2})}{-2\sqrt{1-a^{2}-e^{2}}}\\
A_{-}^{eqKN}  & =\frac{e^{2}(\Phi-a)}{-2\sqrt{1-a^{2}-e^{2}}}%
+\frac{+2a(1-\sqrt{1-a^{2}-e^{2}})-\Phi(a^{2}+e^{2})}{-2\sqrt{1-a^{2}%
-e^{2}}}%
\end{align}
In going from $(\ref{arxidfle})$ to $(\ref{finadefle})$ we used the identity
proven in \cite{KraniotisLightI} $($eqn.$(52)).$ The tuples of numbers for the
beta parameters of the generalized hypergeometric functions in Equation
$($\ref{finadefle}$)$ are defined in Equation $(53)$ of
\cite{KraniotisGVlightII}. Also we defined:%
\begin{equation}
\mathbf{z}_{\pm}^{\mathbf{r}}=\left(  \frac{1}{\omega},\kappa_{\pm}^{\prime
2},\mu^{\prime2}\right)  ,\mathbf{z}_{\mathbf{A}}^{\mathbf{r}}=\left(
\frac{1}{\omega},\mu^{\prime2}\right)  .
\end{equation}

The angle of deflection $\delta$ of light rays from the
gravitational field of a galactic electrically charged rotating
black hole or a massive charged rotating star is defined to be the
deviation of $\Delta\phi_{eKN}^{GTR}$ from $\pi$
\begin{equation}
\fbox{$\delta_{eKN}:=\Delta\phi_{eKN}^{GTR}-\pi.\label{ISimerDelfKN}$}%
\end{equation}
The four roots of the quartic $($\ref{quarweier}$)$, in terms of which the
variables of the generalized hypergeometric functions of Appell and
Lauricella's are written in Equation $(\ref{finadefle}),$ can be expressed in
a very elegant compact closed analytic form in terms of the
Weierstra\ss \ function $\wp(x,g_{2},g_{3})$ and its derivative. Here, we
present the formulae and their derivation is relegated to the appendix A.%
\begin{align}
\alpha & =\frac{1}{2}\frac{\wp^{\prime}(-x_{1}/2+\omega)-\wp^{\prime}(x_{1}%
)}{\wp(-x_{1}/2+\omega)-\wp(x_{1})},\label{maxweierstrass}\\
\beta & =\frac{1}{2}\frac{\wp^{\prime}(-x_{1}/2+\omega+\omega^{\prime}%
)-\wp^{\prime}(x_{1})}{\wp(-x_{1}/2+\omega+\omega^{\prime})-\wp(x_{1})},\label{weirg3}\\
\gamma & =\frac{1}{2}\frac{\wp^{\prime}(-x_{1}/2+\omega^{\prime})-\wp^{\prime
}(x_{1})}{\wp(-x_{1}/2+\omega^{\prime})-\wp(x_{1})}\label{weirstathec},\\
\delta & =\frac{1}{2}\frac{\wp^{\prime}(-x_{1}/2)-\wp^{\prime}(x_{1})}%
{\wp(-x_{1}/2)-\wp(x_{1})}\label{fourth},
\end{align}
where the point $x_{1}$ is defined by the equation:
\begin{equation}
a^{2}-\Phi^{2}=-6\wp(x_{1}),
\end{equation}
and $\omega,\omega^{\prime}$ denotes the half-periods of the
elliptic function $\wp.$ The equations
\begin{equation}
2(a-\Phi)^{2}=4\wp^{\prime}(x_1),-3\wp^2(x_1)+g_2=-e^2(\Phi-a)^2
\end{equation}
determine the Weierstra\ss \ invariants $(g_2,g_3)$ with the
result:
\begin{align}
g_{2}  & =\frac{1}{12}(a^{2}-\Phi^{2})^{2}-e^{2}(\Phi-a)^{2},\\
g_{3}  & =-\frac{1}{216}(a^{2}-\Phi^{2})^{3}-\frac{1}{4}(a-\Phi)^{4}%
-e^{2}(\Phi-a)^{2}\left(  \frac{a^{2}-\Phi^{2}}{6}\right)  .
\end{align}
We are working with dimensionless parameters, effectively setting $M=1$.

\begin{figure}
[ptbh]
\psfrag{fitapanathinaikosgtthirteen}{$\Phi=10,a=0.52$}
\psfrag{e}{$e$} \psfrag{dfgtr}{$\Delta\phi^{GTR}-\pi[{\rm arcs}]$}
\begin{center}
\includegraphics[height=2.4526in, width=3.3797in ]{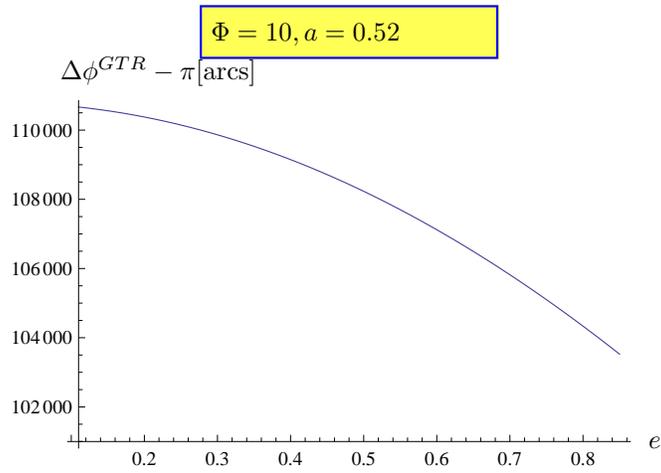}
 \caption{Plot of the deflection angle
$\delta_{eKN}$ versus the electric
charge $|e|,$ for fixed parameters $\Phi=10,a=0.52.$}%
\label{KNeP10A052}%
\end{center}
\end{figure}

\begin{figure}
[ptbh]
\psfrag{trifillaraolepaodekatriathyra13}{$|e|=0.85,a=0.52$}
\psfrag{donner}{$\Delta\phi^{GTR}-\pi[{\rm arcs}]$}
\psfrag{ifi}{$\Phi$}
\begin{center}
\includegraphics[height=2.4526in, width=3.3797in ]{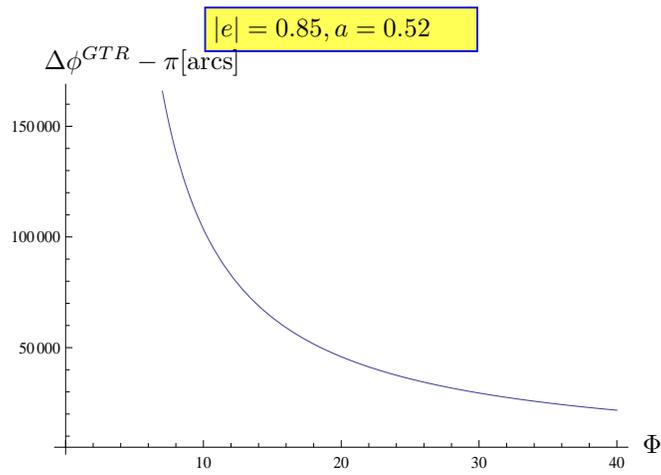}
\caption{Plot
of the deflection angle $\delta_{eKN}$ versus the impact factor
$\Phi,$ for fixed parameters $|e|=0.85,a=0.52.$}%
\label{KNpe05a052}%
\end{center}
\end{figure}

\begin{figure}
[ptbh]
\psfrag{deleknaphiec052}{$\delta_{eKN},a=0.52$}
\psfrag{e}{$e$} \psfrag{phi}{$\Phi$}
\begin{center}
\includegraphics[height=2.9386in, width=4.4244in ]{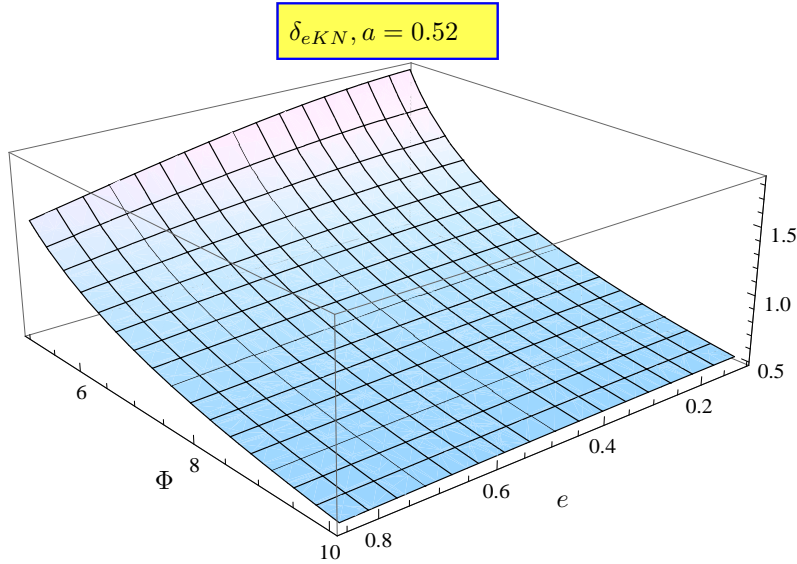}
 \caption{Plot
of the deflection angle $\delta_{eKN}$ as a function of the
parameters $\Phi,e$ for fixed Kerr parameter $a=0.52.$}%
\label{KN3DePa052}%
\end{center}
\end{figure}

\begin{figure}
[ptb]
\psfrag{e}{$e$} \psfrag{pfi}{$\Phi$}
\begin{center}
\includegraphics[height=3.1194in, width=4.1148in ]{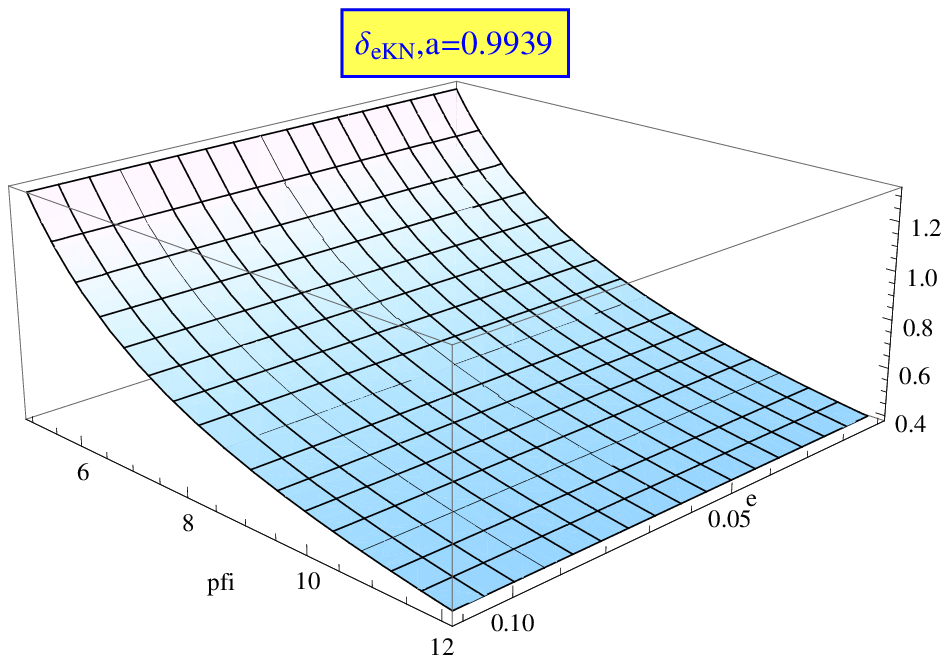}
\caption{Plot
of the deflection angle $\delta_{eKN}$ as a function of the
parameters $\Phi,e$ for fixed Kerr parameter $a=0.9939.$}%
\label{KNBH09939eP}%
\end{center}
\end{figure}

\begin{figure}
[ptb]
\psfrag{geosidirodromos1966electron13georgebill}{$e=0.11,a=0.9939$}
\psfrag{fili}{$\Phi$} \psfrag{spin}{$\Delta\phi^{GTR}-\pi[{\rm arcs}]$}
\begin{center}
\includegraphics[height=2.4526in, width=3.0797in ]{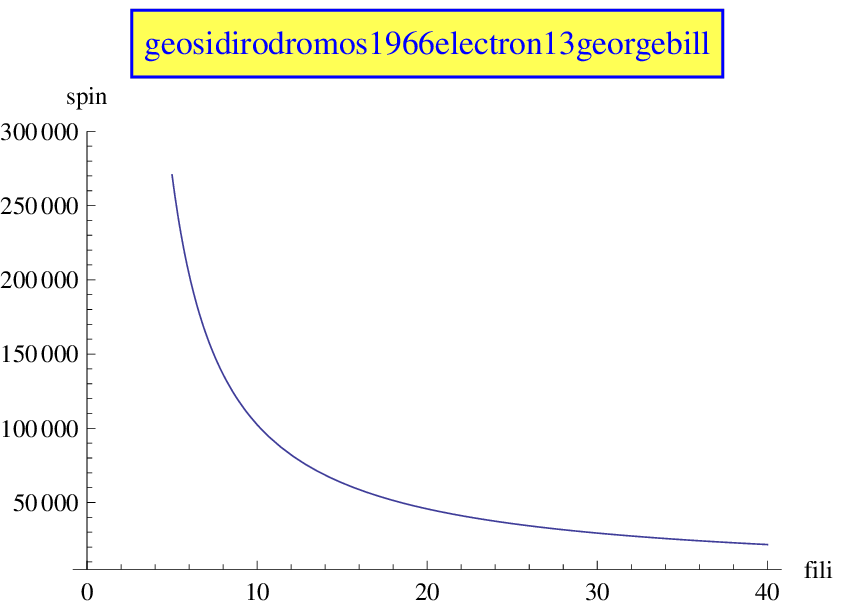}
\caption{Plot of the deflection angle $\delta_{eKN}$ versus the
impact factor
$\Phi,$ for fixed parameters $|e|=0.11,a=0.9939.$}%
\label{KNa09939e011}%
\end{center}
\end{figure}

\begin{figure}
[ptb]
\begin{center}
\includegraphics[height=3.2456in, width=5.1932in ]{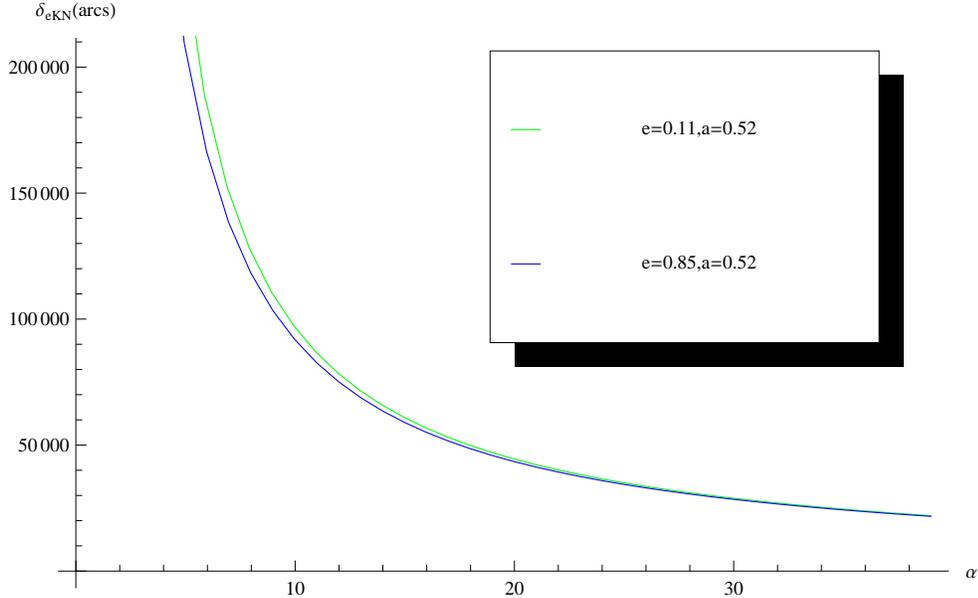}
\caption{The deflection angle $\delta_{eKN}$ versus the root
$\alpha$ for fixed Kerr parameter $a=0.52$ for two different
values of the electric charge.
The values of the electric charge are in units of $\sqrt{G}M_{BH}$ .}%
\label{Physscren052}%
\end{center}
\end{figure}

\begin{figure}
[ptb]
\begin{center}
\includegraphics[height=3.0009in,width=4.7962in]{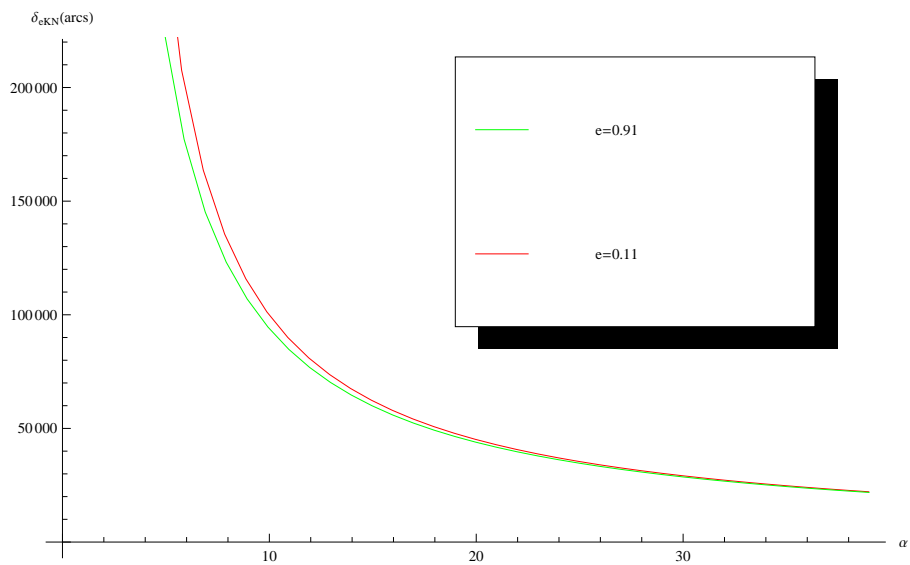}
\caption{The
deflection angle $\delta_{eKN}$ versus the root $\alpha$ for a
fixed Kerr parameter $a=0.26$ for two different values of the
electric charge.
The values of the electric charge are in units of $\sqrt{G}M_{BH}$ .}%
\label{KNedepa026}%
\end{center}
\end{figure}

We computed using our exact analytic formula for the deflection angle,
eqn.(\ref{ISimerDelfKN}), the gravitational bending of light of an unbound
equatorial ray in the gravitational field of a galactic KN-black hole, for
different choices of the spin and electric charge of the black hole and the
impact factor $\Phi.$ We display our results in Tables \ref{Electriccharge04},
\ref{ElectChar011a09939}, \ref{ECH091A026} and Figures \ref{KNeP10A052},
\ref{KNpe05a052}, \ref{KN3DePa052}, \ref{KNBH09939eP}, \ref{KNa09939e011},
\ref{Physscren052}, \ref{KNedepa026}. The values for the Kerr parameter in
Tables \ref{Electriccharge04}, \ref{ElectChar011a09939}, as we mentioned in
the introduction, are in accordance with the central values reported for the
spin of the Galactic centre black hole SgrA* from observation of near infrared
periodic flares \cite{Genzel} and X-ray flares \cite{Aschenbach}\ .

\begin{table}[ht] \centering
\begin{tabular}
[c]{|l|l|}\hline
\textbf{Parameters} & \textbf{Predicted deflection}\\\hline
$a_{Gal}=0.52,e=0.4,\Phi=5$ & $\delta_{eKN}=1.7161=98.3254%
\operatorname{{{}^\circ}}%
=353971\mathrm{\operatorname{arcsec}}$\\
$a_{Gal}=0.52,e=0.4,\Phi=10$ & $\delta_{eKN}=0.529162=30.3188%
\operatorname{{{}^\circ}}%
=109148\operatorname{arcsec}$\\
$a_{Gal}=0.52,e=0.4,\Phi=15$ & $\delta_{eKN}=0.317395=18.1854%
\operatorname{{{}^\circ}}%
=65467.3\operatorname{arcsec}$\\\hline
$a_{Gal}=0.52,e=0.4,\Phi=20$ & $\delta_{eKN}=0.226997=13.006%
\operatorname{{{}^\circ}}%
=46821.5\operatorname{arcsec}$\\\hline
$a_{Gal}=0.52,e=0.4,\Phi=40$ & $\delta_{eKN}=0.106251=6.08775%
\operatorname{{{}^\circ}}%
=21915.9\operatorname{arcsec}$\\\hline
\end{tabular}
\caption{Predictions for light deflection from a galactic electrically
charged rotating black hole with Kerr parameter
$a_{Gal}=0.52\frac{GM_{\rm{BH}}}{c^2}$ and the electric charge parameter
$e=0.4$. The values of the impact parameter $\Phi$ are in units of
$\frac{GM_{\rm{BH}}}{c^2}.$ }\label{Electriccharge04}%
\end{table}%
%

\begin{table}[tbp] \centering
\begin{tabular}
[c]{|l|l|}\hline
\textbf{Parameters} & \textbf{Predicted deflection}\\\hline
$a_{Gal}=0.9939,e=0.11,\Phi=5$ & $\delta_{eKN}=1.3125=75.2005%
\operatorname{{{}^\circ}}%
=270722\operatorname{arcsec}$\\
$a_{Gal}=0.9939,e=0.11,\Phi=10$ & $\delta_{eKN}=0.496749=28.4616%
\operatorname{{{}^\circ}}%
=102461.7\operatorname{arcsec}$\\
$a_{Gal}=0.9939,e=0.11,\Phi=15$ & $\delta_{eKN}=0.306428=17.5571%
\operatorname{{{}^\circ}}%
=63205.4\operatorname{arcsec}$\\\hline
$a_{Gal}=0.9939,e=0.11,\Phi=20$ & $\delta_{eKN}=0.221555=12.6942%
\operatorname{{{}^\circ}}%
=45699.02\operatorname{arcsec}$\\\hline
$a_{Gal}=0.9939,e=0.11,\Phi=40$ & $\delta_{eKN}=0.105111=6.02241%
\operatorname{{{}^\circ}}%
=21680.7\operatorname{arcsec}$\\\hline
\end{tabular}
\caption{Predictions for light deflection from a galactic
electrically charged rotating black hole with Kerr parameter
$a_{Gal}=0.9939\frac{GM_{\rm{BH}}}{c^2}$ and the electric charge
parameter $|e|/\sqrt{G}M_{\rm{BH}}=0.11$. The values of the impact
parameter $\Phi$ are in units of
$\frac{GM_{\rm{BH}}}{c^2}.$ }\label{ElectChar011a09939}%
\end{table}%
%

\begin{table}[tbp] \centering
\begin{tabular}
[c]{|l|l|}\hline
\textbf{Parameters} & \textbf{Predicted deflection}\\\hline
$a_{Gal}=0.26,e=0.91,\Phi=5$ & $\delta_{eKN}=1.57038=89.9759%
\operatorname{{{}^\circ}}%
=323913\operatorname{arcsec}$\\
$a_{Gal}=0.26,e=0.91,\Phi=10$ & $\delta_{eKN}=0.518656=29.7168%
\operatorname{{{}^\circ}}%
=106980\operatorname{arcsec}$\\
$a_{Gal}=0.26,e=0.91,\Phi=15$ & $\delta_{eKN}=0.313795=17.9791%
\operatorname{{{}^\circ}}%
=64724.8\operatorname{arcsec}$\\\hline
$a_{Gal}=0.26,e=0.91,\Phi=20$ & $\delta_{eKN}=0.225192=12.9026%
\operatorname{{{}^\circ}}%
=46449.3\operatorname{arcsec}$\\\hline
$a_{Gal}=0.26,e=0.91,\Phi=40$ & $\delta_{eKN}=0.105866=6.06567%
\operatorname{{{}^\circ}}%
=21836.4\operatorname{arcsec}$\\\hline
\end{tabular}
\caption{Predictions for light deflection from a galactic
electrically charged rotating black hole with Kerr parameter
$a_{Gal}=0.26\frac{GM_{\rm{BH}}}{c^2}$ and the electric charge
parameter $|e|/\sqrt{G}M_{\rm{BH}}=0.91$. The values of the impact
parameter $\Phi$ are in units of
$\frac{GM_{\rm{BH}}}{c^2}.$}\label{ECH091A026}%
\end{table}%
%

%

In Figures \ref{KN3DePa052},\ref{KNBH09939eP} we display 3-d plots of the
deflection angle $\delta_{eKN}$ as a function of the impact factor and the
electric charge, for two fixed values of the Kerr parameter. From these plots
we observe that the smaller the Kerr parameter the larger the deflection, for
fixed values of the parameters $\Phi,e.$ We also observe, the strong
dependence of the deflection angle on the electric charge, for smaller values
of the spin of the black hole, particularly for small values of the impact
factor parameter $\Phi.$%

%


In figures \ref{Physscren052},\ref{KNedepa026} we plot the deflection angle
$\delta_{eKN}$ versus the maximal root $\alpha$-see equation $($%
\ref{maxweierstrass}$)$-of the quartic for fixed Kerr (spin) parameter for two
different values of the electric charge. In figure \ref{Physscren052}, we fix
the Kerr parameter to the value $a=0.52$ and the two different choices for
electric charge: $e=0.11,e=0.85.$ We observe from the analysis that for a
fixed small distance $\alpha$ there is a strong dependence of the deflection
angle on the electric charge the black hole carries: the larger the electric
charge $e,$ the smaller $\delta_{eKN}.$

In figure \ref{Physscren052}, the values of the electric charge, for the SgrA*
galactic black hole correspond to:%
\begin{align}
e  &
=0.85\sqrt{6.6743\times10^{-8}}4.06\times10^{6}\times1.9884\times
10^{33}\mathrm{esu=}1.77\times10^{36}\mathrm{esu}\Leftrightarrow
5.94\times10^{26}%
\operatorname{C}%
,\nonumber\\
e  & =0.11\sqrt{6.6743\times10^{-8}}4.06\times10^{6}\times1.9884\times
10^{33}\mathrm{esu=}2.29\times10^{35}\mathrm{esu}\Leftrightarrow
7.65\times10^{25}%
\operatorname{C}%
.\nonumber\\
&
\end{align}
Concerning the tentative values for the electric charge $e$ we
used in applying our exact solutions for the case of SgrA* black
hole we note that their likelihood is debatable: There is an expectation that the
electric charge trapped in the galactic nucleous will not likely
reach so high values as the ones close to the extremal values
predicted in (\ref{FundEcGMa}) that allow the avoidance of a naked
singularity. However, more precise statements on the electric
charge's magnitude of the galactic black hole or its upper bound
will only be reached once the relativistic effects predicted in
this work are measured and a comparison of the theory we developed
with experimental data will take place \footnote{In this regard,
we also mention that the author in \cite{LIORIO}, under the
assumption that the curved geometry surrounding the massive object
in the Galactic Centre is a Reissner-Nordstr\"{o}m (RN) spacetime,
obtained an upper bound of $e\lesssim 3.6\times10^{27}$C. This
upper bound does not distinguish yet between a RN black hole
scenario and a RN naked singularity scenario.}.

%

\begin{subappendices}%

\subsection{Exact solution for the periastron advance in Kerr-Newman
spacetime} \label{SPEriastronKN}

In this appendix of the section we derive the closed form solution for the
periastron advance for a timelike equatorial non-circular orbit and apply it
to the computation of this relativistic effect for the observed orbits of $S-
$stars in the central arecsecond of our Galaxy assuming that the galactic
centre region Sagittarius A* harbours a supermassive rotating black hole in
which a net electric charge has been trapped inside the event horizon $r_{+},
$ to form a charged KN black hole.

The relevant differential equation is given by:
\begin{equation}
\frac{\mathrm{d}\phi}{\mathrm{d}r}=\frac{e^{2}(L-aE)+Lr^{2}-2Mr(L-aE)}{\pm(r^{2}+a^{2}+e^{2}-2Mr)\sqrt{R}},
\end{equation}
where the quartic polynomial is given by
\begin{align}
R  & =[(r^{2}+a^{2})E-aL]^{2}-\Delta_{KN}[r^{2}+(L-aE)^{2}]\nonumber\\
& =r^{4}(E^{2}-1)+2Mr^{3}+r^{2}(a^{2}E^{2}-L^{2}-e^{2}%
-a^{2})\nonumber\\
& +2Mr[L^{2}+a^{2}E^{2}-2aLE]-e^{2}L^{2}-E^{2}e^{2}%
a^{2}+2ae^{2}EL.
\end{align}
Using again the partial fractions technique and performing similar
manipulations as in eqn.(\ref{manipulativeop}), we integrate from
the periastron
distance $r_{P}$ to the apoastron distance $r_{A}$:%
\begin{align}
\Delta\phi_{teKN}^{GTR}  & =\int_{r_{P}}^{r_{A}}\frac{e^{2}(L-aE)+Lr^{2}%
-2Mr(L-aE)}{(r^{2}+a^{2}+e^{2}-2Mr)\sqrt{R}%
}\mathrm{d}r\nonumber\\
& =\int_{r_{P}}^{r_{A}}\frac{L}{\sqrt{R}}\mathrm{d}r+\int_{r_{P}}^{r_{A}}\frac{A_{teKN}^{+}}{(r-r_{+})\sqrt{R}}%
\mathrm{d}r+\int_{r_{P}}^{r_{A}}\frac{A_{teKN}^{-}}{(r-r_{-})\sqrt{R}%
}\mathrm{d}r,
\end{align}
where
\begin{equation}
A_{teKN}^{\pm}=\frac{\pm L(a^{2}+e^{2})\mp2aEr_{\pm}^{\prime}\mp e^{2}(L-aE)}%
{-2\sqrt{1-a^{2}-e^{2}}}.
\end{equation}

Applying the transformation:%
\begin{equation}
z=\frac{1}{\omega}\frac{r^{\prime}-\alpha_{\mu+1}}{r^{\prime}-\alpha_{\mu+2}%
}=\frac{\alpha-\gamma}{\alpha-\beta}\frac{r^{\prime}-\beta}{r^{\prime}-\gamma}%
\end{equation}
and organizing the roots of the radial polynomial and the radii of the event
and Cauchy horizon in the ascending order of magnitude%
\begin{equation}
\alpha_{\rho}>\alpha_{\sigma}>\alpha_{\nu}>\alpha_{i},
\end{equation}
with the correspondense $\alpha_{\rho}=\alpha_{\mu}=\alpha,\alpha_{\sigma
}=\alpha_{\mu+1}=\beta,\alpha_{\nu}=\alpha_{\mu+2}=\gamma,\alpha_{i}%
=\alpha_{\mu-i},i=1,2,3,\alpha_{\mu-1}=a_{\mu-2}=r_{\pm}^{\prime},\alpha
_{\mu-3}=\delta$ we compute the exact analytic result in terms of Appell's
hypergeometric function $F_{1}$ and Gau\ss 's ordinary hypergeometric
function:%
\begin{align}
\Delta\phi_{teKN}^{GTR}  & =2%
\Biggl[%
-\frac{\omega^{3/2}}{H_{+}}A_{teKN}^{+}F_{1}\left(  \frac{3}{2},1,\frac{1}%
{2},2,\kappa_{+}^{t2},\kappa^{\prime2}\right)  \frac{\Gamma\left(  \frac{3}%
{2}\right)  \Gamma\left(  \frac{1}{2}\right)  }{\Gamma(2)}\nonumber\\
& +\frac{\sqrt{\omega}}{H_{+}}A_{teKN}^{+}F_{1}\left(  \frac{1}{2}%
,1,\frac{1}{2},1,\kappa_{+}^{t2},\kappa^{\prime2}\right)  \frac{\Gamma
^{2}\left(  \frac{1}{2}\right)  }{\Gamma(1)}\nonumber\\
& -\frac{\omega^{3/2}}{H_{-}}A_{teKN}^{-}F_{1}\left(  \frac{3}{2},1,\frac
{1}{2},2,\kappa_{-}^{t2},\kappa^{\prime2}\right)  \frac{\Gamma\left(  \frac
{3}{2}\right)  \Gamma\left(  \frac{1}{2}\right)  }{\Gamma(2)}\nonumber\\
& +\frac{\sqrt{\omega}}{H_{-}}A_{teKN}^{-}F_{1}\left(  \frac{1}{2}%
,1,\frac{1}{2},1,\kappa_{-}^{t2},\kappa^{\prime2}\right)  \frac{\Gamma
^{2}\left(  \frac{1}{2}\right)  }{\Gamma(1)}%
\Biggr]%
\nonumber\\
& +\frac{2\sqrt{\omega}L}{\sqrt{(1-E^{2})(\alpha-\beta)(\beta-\delta)}%
}F\left(  \frac{1}{2},\frac{1}{2},1,\kappa^{\prime2}\right)  \pi\nonumber\\
& \label{RelativiPeriastronPrec}%
\end{align}

The variables of the hypergeometric functions are given in terms
of the roots
of the quartic and the radii of the horizons by the expressions:%
\begin{equation}
\kappa_{\pm}^{t2}:=\frac{\alpha-\beta}{\alpha-\gamma}\frac{r_{\pm}-\gamma
}{r_{\pm}-\beta},\text{ \ }\kappa^{\prime2}:=\frac{\alpha-\beta}{\alpha
-\gamma}\frac{\delta-\gamma}{\delta-\beta}.
\end{equation}
The periapsis advance for an equatorial non-circular timelike geodesic in
Kerr-Newman spacetime is defined as follows:%
\begin{equation}
\fbox{$\delta_{P}^{teKN}:=\Delta\phi_{teKN}^{GTR}-2\pi\label{advaadva}$}%
\end{equation}
We applied our closed form formula $(\ref{RelativiPeriastronPrec}%
),(\ref{advaadva})$ for calculating the relativistic periapsis
advance for the observed orbits of $S$-stars in the central
arcsecond of the Milky Way. By doing this exercise, we gain an
appreciation of the effect of the electric charge of the rotating
galactic black hole (we assume that the KN solution describes the
curved spacetime geometry around SgrA*) on this observable. We
also assume that the angular momentum axis of the orbit is
co-aligned with the spin axis of the black hole and that the
$S-$stars can be treated as neutral test particles. Indeed, we
present the results of our computations in Tables
\ref{StarS2eaSgrA*},\ref{AstroS14eaSgraA*}, that correspond to the
orbits of the stars $S2$ and $S14$ respectively\footnote{The
parameters are consistent with data for the periastron, apoastron
distances and orbital period for the stars $S2,S14$
\cite{Eisenhauer} (see also \cite{GeorgeVKraniotis}).}. \ For
fixed values of the parameters $L,E,a $ we \ calculated the
periapsis advance for different values of the electric charge. We
observe the significant contribution of the electric charge on the
phenomenon of periapsis advance in the theory of general
relativity. Varying the electric charge in the range
$(0.1-0.85)\sqrt{G}M_{BH}$ we see that the effect due to the
electric
charge on the periastron advance is of the order of $81.4\operatorname{arcsec}%
/rev.$ for the star $S2,$with the observation: the larger the electric charge
the smaller the periapsis advance. Similar results hold for the star $S14$
where the effect due to the parameter $e$ is computed to be of the order of
$87\operatorname{arcsec}/rev$ as $e$ varies in the range $(0.1-0.85)\sqrt%
{G}M_{BH}.$ A more precise analysis would involve the calculation
of relativistic periapsis advance for general timelike orbits,
polar or inclined non-equatorial in the KN(a)dS spacetime, which
is beyond the scope of the current publication \footnote{Such an
analysis will be a subject of a future publication}. However,
since the effect due to the parameter $e$ is significant already
at this level of analysis and since we are entering an era of
precision in observational astronomy the effect due to the
electric charge of the KN spacetime singularity should be taken
into account in the interpretations of
future measurements for the relativistic effect of periapsis advance \cite{Andrea}.%

\begin{table}[tbp] \centering
\begin{tabular}
[c]{|l|l|}\hline
\textbf{Parameters for the star }$S2$ & \textbf{Periapsis advance}\\\hline
$a=0.52,e=0.1,L=75.4539876,E=0.999979485$ & $\delta_{P}^{teKN}=676.5\frac
{\operatorname{arcsec}}{revol.}$\\
$a=0.52,e=0.33,L=75.4539876,E=0.999979485$ & $\delta_{P}^{teKN}=665.2\frac
{\operatorname{arcsec}}{revol.}$\\
$a=0.52,e=0.85,L=75.4539876,E=0.999979485$ & $\delta_{P}^{teKN}=595.1\frac
{\operatorname{arcsec}}{revol.}$\\\hline
\end{tabular}
\caption{Periastron precession for the star $S2$ in the central arcsecond of
the galactic centre, using the exact formula
$(\ref{RelativiPeriastronPrec})$, for three different values of the electric
charge of the galactic black hole.The Kerr parameter is
$a_{Gal}=0.52\frac{GM_{\rm BH} }{c^2}$. We assume a central black hole mass
$M_{\rm BH}=4.06\times 10^6M_{\odot}$}\label{StarS2eaSgrA*}%
\end{table}%
%

\begin{table}[tbp] \centering
\begin{tabular}
[c]{|l|l|}\hline
\textbf{Parameters for the star }$S14$ & \textbf{Periapsis advance}\\\hline
$a=0.52,e=0.11,L=72.9456205,E=0.999988863$ & $\delta_{P}^{teKN}=723.432\frac
{\operatorname{arcsec}}{revol.}$\\
$a=0.52,e=0.33,L=72.9456205,E=0.999988863$ & $\delta_{P}^{teKN}=711.595\frac
{\operatorname{arcsec}}{revol.}$\\
$a=0.52,e=0.85,L=72.9456205,E=0.999988863$ & $\delta_{P}^{teKN}=636.568\frac
{\operatorname{arcsec}}{revol.}$\\\hline
\end{tabular}
\caption{Periastron precession for the star $S14$ in the central arcsecond
of the galactic centre, using the exact formula
$(\ref{RelativiPeriastronPrec})$, for three different values of the electric
charge of the galactic black hole.The Kerr parameter is
$a_{Gal}=0.52\frac{GM_{\rm BH} }{c^2}$. We assume a central black hole mass
$M_{\rm BH}=4.06\times 10^6M_{\odot}$}\label{AstroS14eaSgraA*}%
\end{table}%

\end{subappendices}%

\section{Exact solution for the deflection angle of unbound equatorial orbits
in Kerr-Newman-de Sitter spacetime \label{defllambda}}

Assume that $\Lambda>0.$ Then the relevant differential equation for the exact
computation of the deflection angle of an equatorial ray in the field of an
electrically charged rotating black hole with a positive cosmological constant
is given by%
\begin{equation}
\mathrm{d}\phi=\Xi^{2}(\Phi-a)\frac{\mathrm{d}r}{\sqrt{R^{\prime}}}%
+\frac{a\Xi^{2}}{\Delta_{r}^{KN}}\frac{[(r^{2}+a^{2})-a\Phi]}{\sqrt%
{R^{\prime}}}.
\end{equation}

Using \ the partial fractions technique for the second term we write:%
\begin{equation}
\frac{a\Xi^{2}}{\Delta_{r}^{KN}}\frac{[(r^{2}+a^{2})-a\Phi]}{\sqrt%
{R^{\prime}}}=\frac{A^{1}}{r-r_{\Lambda}^{+}}+\frac{A^{2}}{r-r_{\Lambda}^{-}%
}+\frac{A^{3}}{r-r_{+}}+\frac{A^{4}}{r-r_{-}}%
\end{equation}
where $r_{\Lambda}^{\pm},r_{\pm},$ are the four real roots of
$\Delta_{r}^{KN}.$ Thus one of the integrals we need to calculate is:%
\begin{equation}
I_{\Lambda}^{1}=\frac{1}{\sqrt{\Xi^{2}\left(  1+\frac{\Lambda}{3}%
(\Phi-a)^{2}\right)  }}\int_{\alpha}^{r_{\Lambda}^{+}/2}\frac{A^{1}%
\mathrm{d}r}{(r-r_{\Lambda}^{+})\sqrt{(r-\alpha)(r-\beta)(r-\gamma
)(r-\delta)}}%
\end{equation}
Indeed, we compute in closed analytic form:%
\begin{equation}
I_{\Lambda}^{1}=-\frac{A^{1}}{\sqrt{\Xi^{2}\left(  1+\frac{\Lambda}{3}%
(\Phi-a)^{2}\right)  }}\frac{\rho_{1}\omega^{\prime}}{\sqrt{\rho_{1}}%
}H_{\Lambda e}^{+}F_{D}\left(  \frac{1}{2},\mathbf{\beta}_{4}^{\Lambda3}%
,\frac{3}{2},\mathbf{z}_{\mathbf{\Lambda}^{+}}^{\mathbf{r}}\right)
\frac{\Gamma(1/2)}{\Gamma(3/2)}.
\end{equation}
The tuple of variables for the Lauricella's fourth hypergeometric function
$F_{D}$ is defined in terms of the horizons and the radial roots of the
Kerr-Newman-de Sitter black hole as follows:%
\begin{equation}
\mathbf{z}_{\mathbf{\Lambda}^{+}}^{\mathbf{r}}:=\left(  \frac{r_{\Lambda}%
^{+}-2\alpha}{r_{\Lambda}^{+}-2\beta},\frac{\beta-\gamma}{\alpha-\gamma}%
\frac{r_{\Lambda}^{+}-2\alpha}{r_{\Lambda}^{+}-2\beta},\frac{\beta-\delta
}{\alpha-\delta}\frac{r_{\Lambda}^{+}-2\alpha}{r_{\Lambda}^{+}-2\beta}%
,\frac{r_{\Lambda}^{+}-\beta}{r_{\Lambda}^{+}-\alpha}\frac{r_{\Lambda}%
^{+}-2\alpha}{r_{\Lambda}^{+}-2\beta}\right)  ,
\end{equation}
while we also define:%
\begin{equation}
\rho_{1}:=\frac{r_{\Lambda}^{+}-\beta}{r_{\Lambda}^{+}-\alpha}\frac
{r_{\Lambda}^{+}-2\alpha}{r_{\Lambda}^{+}-2\beta}%
\end{equation}
and%
\begin{equation}
H_{\Lambda e}^{+}:=\frac{\alpha-\beta}{|\beta-\alpha|}\frac{1}{r_{\Lambda}%
^{+}-\beta}\frac{1}{\sqrt{\omega^{\prime}(\gamma-\alpha)(\delta-\alpha)}}.
\end{equation}

In addition, from the first term we compute exactly in terms of
Appell's hypergeometric function:%
\begin{align}
& \Xi^{2}(\Phi-a)\int_{\alpha}^{r_{\Lambda}^{+}/2}\frac{\mathrm{d}r}%
{\sqrt{(r-\alpha)(r-\beta)(r-\gamma)(r-\delta)}}\nonumber\\
& =\frac{\Xi^{2}(\Phi-a)}{\sqrt{\Xi^{2}\left(  1+\frac{\Lambda}{3}%
(\Phi-a)^{2}\right)  }}\frac{\rho_{1}\omega^{\prime}}{\sqrt{\rho_{1}%
\omega^{\prime}(\gamma-\alpha)(\delta-\alpha)}}\frac{\Gamma(1/2)}{\Gamma
(3/2)}F_{1}\left(  \frac{1}{2},\mathbf{\beta}_{\mathbf{A}}^{\mathbf{ra}}%
,\frac{3}{2},\mathbf{z}_{\mathbf{A\Lambda}^{+}}^{\mathbf{r}}\right)
\nonumber\\
&
\end{align}
In total we get
\begin{align}
\Delta\phi_{eKN\Lambda}^{GTR}  & =\frac{\Xi^{2}(\Phi-a)}{\sqrt{\Xi
^{2}\left(  1+\frac{\Lambda}{3}(\Phi-a)^{2}\right)  }}\frac{\rho_{1}%
\omega^{\prime}2}{\sqrt{\rho_{1}\omega^{\prime}(\gamma-\alpha
)(\delta-\alpha)}}F_{1}\left(  \frac{1}{2},\mbox{\boldmath${\beta}_{\mathbf{A}%
}$}^{\mathbf{ra}},\frac{3}{2},\mathbf{z}_{\mathbf{A\Lambda}^{+}}^{\mathbf{r}%
}\right) \nonumber\\
& -\frac{A^{1}}{\sqrt{\Xi^{2}\left(  1+\frac{\Lambda}{3}(\Phi
-a)^{2}\right)
}}\frac{\rho_{1}\omega^{\prime}}{\sqrt{\rho_{1}}}H_{\Lambda
e}^{+}F_{D}\left(  \frac{1}{2},\mbox{\boldmath${\beta}_{4}^{\Lambda3}$},\frac{3}%
{2},\mathbf{z}_{\mathbf{\Lambda}^{+}}^{\mathbf{r}}\right)  2\nonumber\\
& -\frac{A^{2}}{\sqrt{\Xi^{2}\left(  1+\frac{\Lambda}{3}(\Phi
-a)^{2}\right)
}}\frac{\rho_{1}\omega^{\prime}}{\sqrt{\rho_{1}}}H_{\Lambda
e}^{-}F_{D}\left(  \frac{1}{2},\mbox{\boldmath${\beta}_{4}^{\Lambda3}$},\frac{3}%
{2},\mathbf{z}_{\mathbf{\Lambda}^{-}}^{\mathbf{r}}\right)  2\nonumber\\
& -\frac{A^{3}}{\sqrt{\Xi^{2}\left(  1+\frac{\Lambda}{3}(\Phi
-a)^{2}\right)
}}\frac{\rho_{1}\omega^{\prime}}{\sqrt{\rho_{1}}}H_{\Lambda
re}^{+}F_{D}\left(  \frac{1}{2},\mbox{\boldmath${\beta}_{4}^{\Lambda3}$},\frac{3}%
{2},\mathbf{z}_{\mathbf{\Lambda r}^{+}}^{\mathbf{r}}\right)  2\nonumber\\
& -\frac{A^{4}}{\sqrt{\Xi^{2}\left(  1+\frac{\Lambda}{3}(\Phi
-a)^{2}\right)
}}\frac{\rho_{1}\omega^{\prime}}{\sqrt{\rho_{1}}}H_{\Lambda
re}^{-}F_{D}\left(  \frac{1}{2},\mbox{\boldmath${\beta}_{4}^{\Lambda3}$},\frac{3}%
{2},\mathbf{z}_{\mathbf{\Lambda r}^{-}}^{\mathbf{r}}\right)  2.
\end{align}
where the tuple for the beta parameters of the Lauricella's hypergeometric
function is defined as follows:%
\begin{equation}
\mbox{\boldmath${\beta}_{4}^{\Lambda3}$}:=\left(  -1,\frac{1}{2},\frac{1}{2},1\right)  .
\end{equation}
In producing the analytic solution we applied the transformations:%
\begin{equation}
z=\frac{1}{\omega^{\prime}}\frac{r-\alpha}{r-\beta},z\rightarrow\rho
_{1}z^{\prime},
\end{equation}
and
\begin{equation}
\omega^{\prime}:=\frac{r_{\Lambda}^{+}-\alpha}{r_{\Lambda}^{+}-\beta}.
\end{equation}
In addition we defined the tuple:
\begin{equation}
\mathbf{z_{A\Lambda^+}^r}:=\left(\frac{r_{\Lambda}^{+}-2\alpha}{r_{\Lambda}^{+}-2\beta}
\frac{\beta-\gamma}{\alpha-\gamma},\frac{r_{\Lambda}^{+}-2\alpha}{r_{\Lambda}^{+}-2\beta}
\frac{\beta-\delta}{\alpha-\delta}\right)
\end{equation}

A complete phenomenological analysis of our exact solutions in the
presence of the cosmological constant $\Lambda$ will be a subject
of a separate publication \cite{KRANIOTISGVQEDPEMBH}.
Nevertheless, it is evident from the closed form solutions we
derived in this work that the cosmological constant \textit{does}
contribute to the gravitational bending of light in the KNdS
spacetime.

\section{ The shadow of the electrically charged rotating
(Kerr-Newman) black hole}

\begin{figure}
[ptb]
\begin{center}
\includegraphics[height=3.7456in, width=5.1932in ]{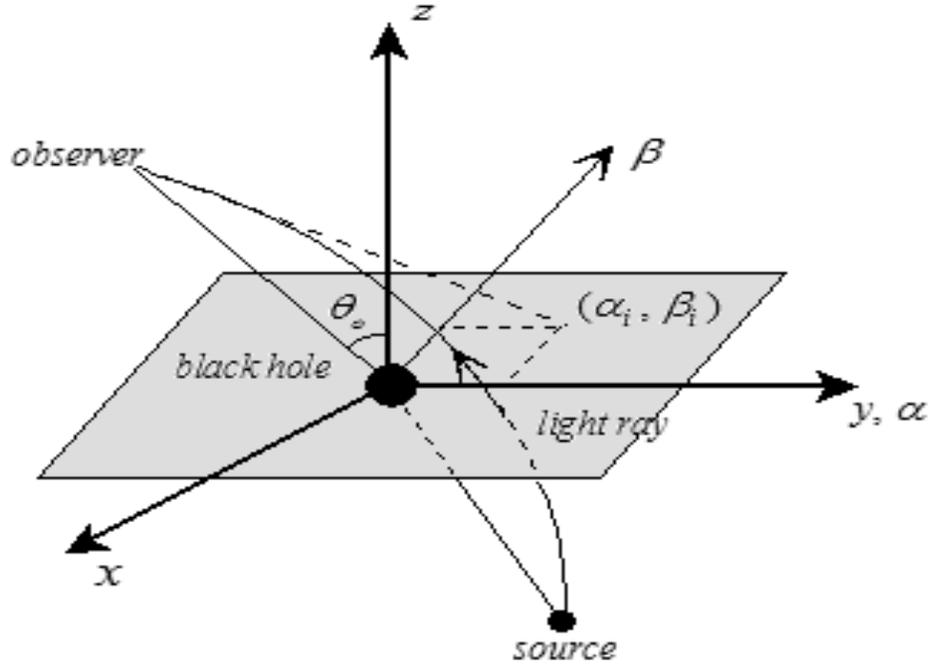}
\caption{The Kerr-Newman black hole gravitational lens geometry. The reference frame is chosen so that, as seen from
infinity, the black hole is rotating around the $z$-axis.}%
\label{KNBHGEOMETRY}%
\end{center}
\end{figure}

The conditions for the spherical photon orbits in Kerr-Newman spacetime as
implemented in section \ref{sppolarkn}, yielded equations $($%
\ref{arxikescondipcc}$)$ for the parameter $\Phi$ and Carter's
constant $\mathcal{Q}.$ These are also the conditions for the
photon to escape at infinity. When we treat the KN black hole as a
gravitational lens, following the procedure developed in
\cite{KraniotisGVlightII} \ for the case of a Kerr gravitational
lens, and assuming large observer's distance $r_{O}($ i.e.
$r_{O}\rightarrow\infty)$ we derive simplified expressions
relating the coordinates \footnote{We also assume without loss of
generality that $\phi_O=0$.}
$(\alpha_{i},\beta_{i})=(-r_{O}^{2}\sin\theta_{O}\frac
{\mathrm{d}\phi}{\mathrm{d}r}|_{r=r_{O}},r_{O}^{2}\frac{\mathrm{d}\theta
}{\mathrm{d}r}|_{r=r_{O}})$ on the observer's image plane (see
figure \ref{KNBHGEOMETRY}) to the integrals of
motion:%
\begin{align}
\Phi & \simeq-\alpha_{i}\sin\theta_{O},\; \mathcal{Q}
\simeq\beta_{i}^{2}+(\alpha_{i}^{2}-a^{2})\cos^{2}\theta _{O}.
\label{observationKNs}%
\end{align}

\begin{figure}
[ptbh]
\psfrag{spin09939charge011observer60}{$a=0.9939,e=0.11,\theta_O=\frac{\pi}{3}$}
\begin{center}
\includegraphics[height=2.3386in, width=2.1386in ]{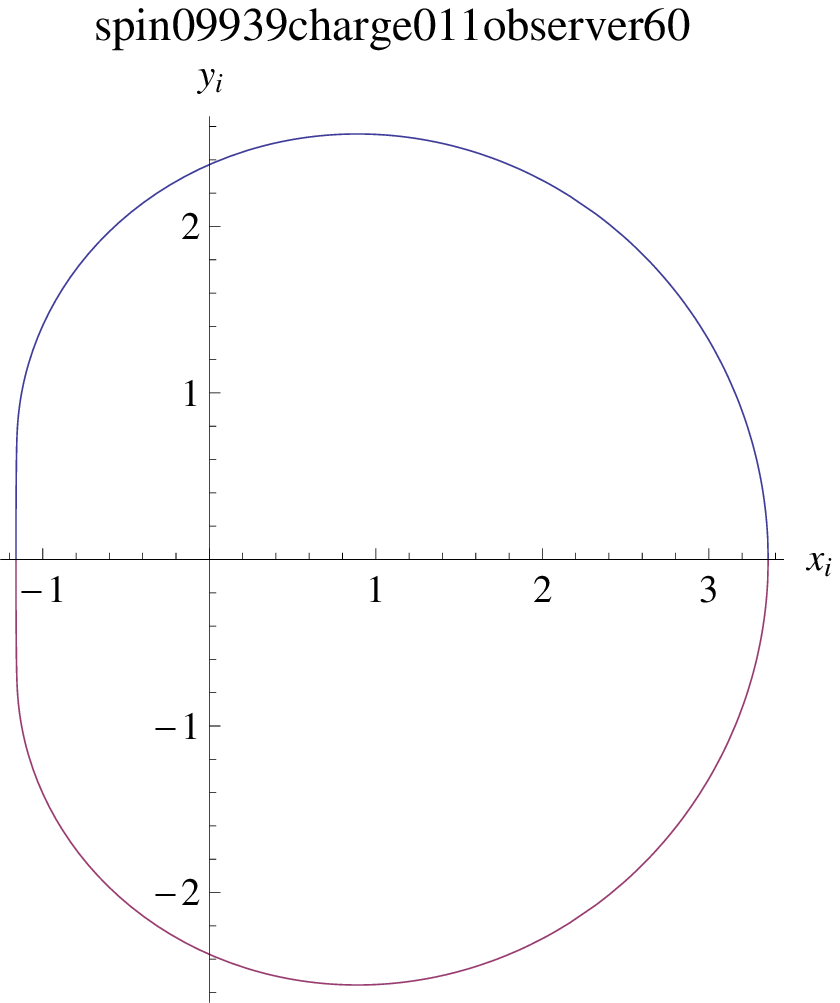}
 \caption{The boundary of the shadow of the charged rotating KN black hole for Kerr parameter
 $a=0.9939$, $e=0.11$ for an observer at polar position $\theta_O=\pi/3$.}%
\label{SWKN993911P3}%
\end{center}
\end{figure}

\begin{figure}
[ptbh]
\psfrag{obsevpi3angmom052eleccharge085}{$a=0.52,e=0.85,\theta_O=\frac{\pi}{3}$}
\begin{center}
\includegraphics[height=2.3386in, width=2.1386in ]{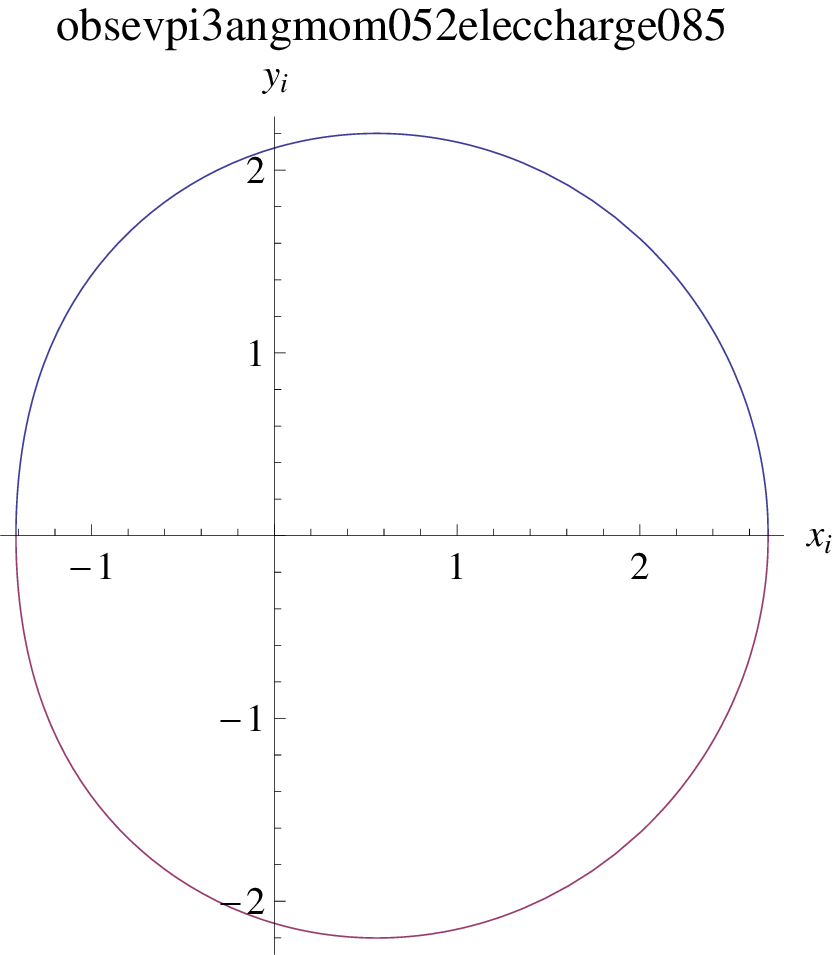}
\caption{The boundary of the shadow of the charged rotating KN black hole for Kerr parameter
 $a=0.52$, $e=0.85$ for an observer at polar position $\theta_O=\pi/3$.}%
\label{SWKN5285PI3}%
\end{center}
\end{figure}

\begin{figure}
[ptbh]
\psfrag{parathr60spin052emcharg04}{$a=0.52,e=0.4,\theta_O=\frac{\pi}{3}$}
\begin{center}
\includegraphics[height=2.3386in, width=2.1386in ]{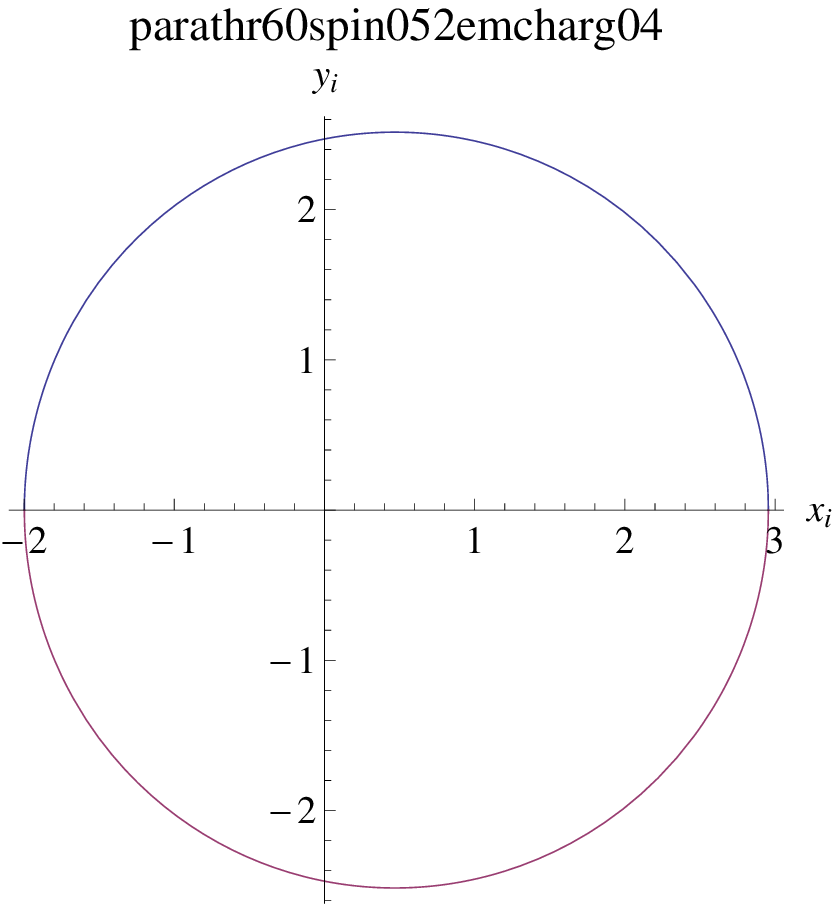}
 \caption{The boundary of the shadow of the charged rotating KN black hole for Kerr parameter
 $a=0.52$, $e=0.4$ for an observer at polar position $\theta_O=\pi/3$.}%
\label{EMROTSWKN5204PI3}%
\end{center}
\end{figure}

\begin{figure}
[ptbh]
\psfrag{obserpol60a052electriccharge011}{$a=0.52,e=0.11,\theta_O=\frac{\pi}{3}$}
\begin{center}
\includegraphics[height=2.3386in, width=2.1386in ]{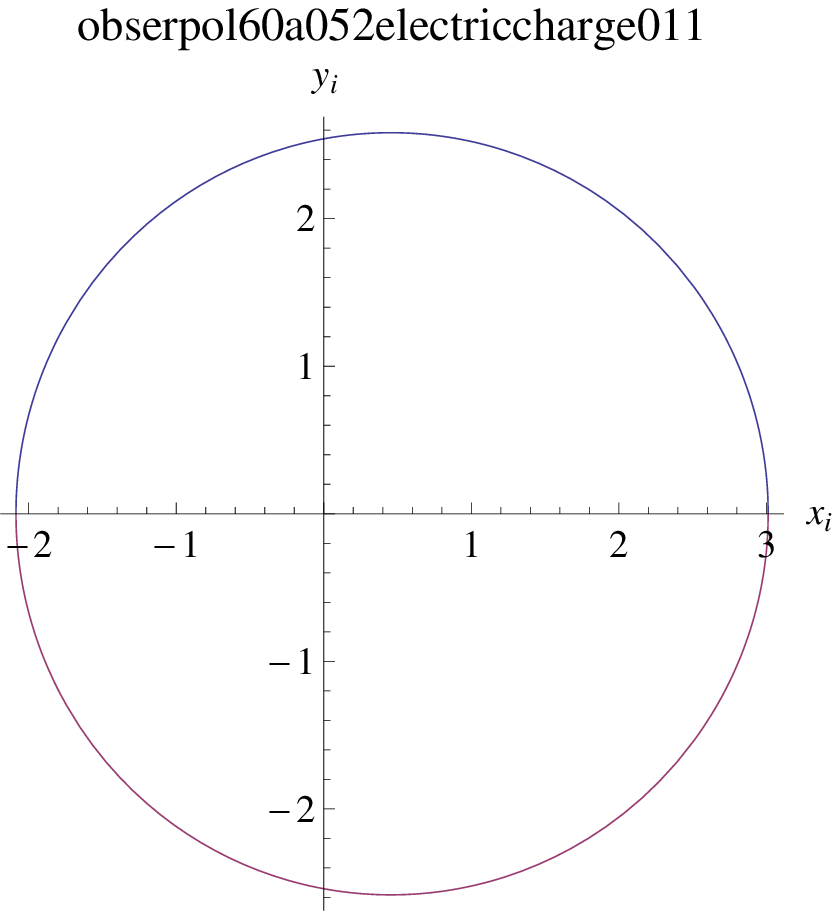}
\caption{The boundary of the shadow of the charged rotating KN black hole for Kerr parameter
 $a=0.52$, $e=0.11$ for an observer at polar position $\theta_O=\pi/3$.}%
\label{EMSWKN5211PI3}%
\end{center}
\end{figure}

\begin{figure}
[ptbh]
\psfrag{obserequatoa09939emcharge011}{$a=0.9939,e=0.11,\theta_O=\frac{\pi}{2}$}
\begin{center}
\includegraphics[height=2.3386in, width=2.1386in ]{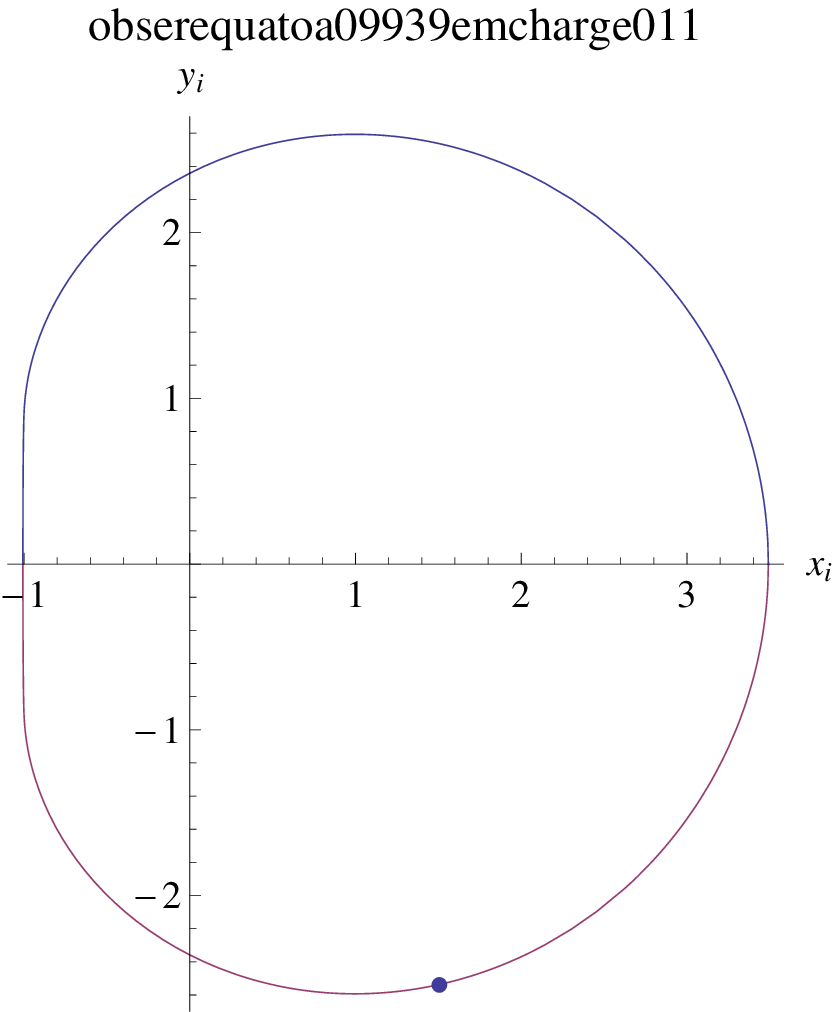}
\caption{The boundary of the shadow of the charged rotating KN black hole for Kerr parameter
 $a=0.9939$, $e=0.11$ for an equatorial observer, i.e. located at polar position $\theta_O=\pi/2$. With the blue dot
 we exhibit the image solution of the KN lens equations, see Table 10.}%
\label{EMSWKN993911PI2}%
\end{center}
\end{figure}

\begin{figure}
[ptbh]
\psfrag{isimeriparatha052emcharge085}{$a=0.52,e=0.85,\theta_O=\frac{\pi}{2}$}
\begin{center}
\includegraphics[height=2.3386in, width=2.1386in ]{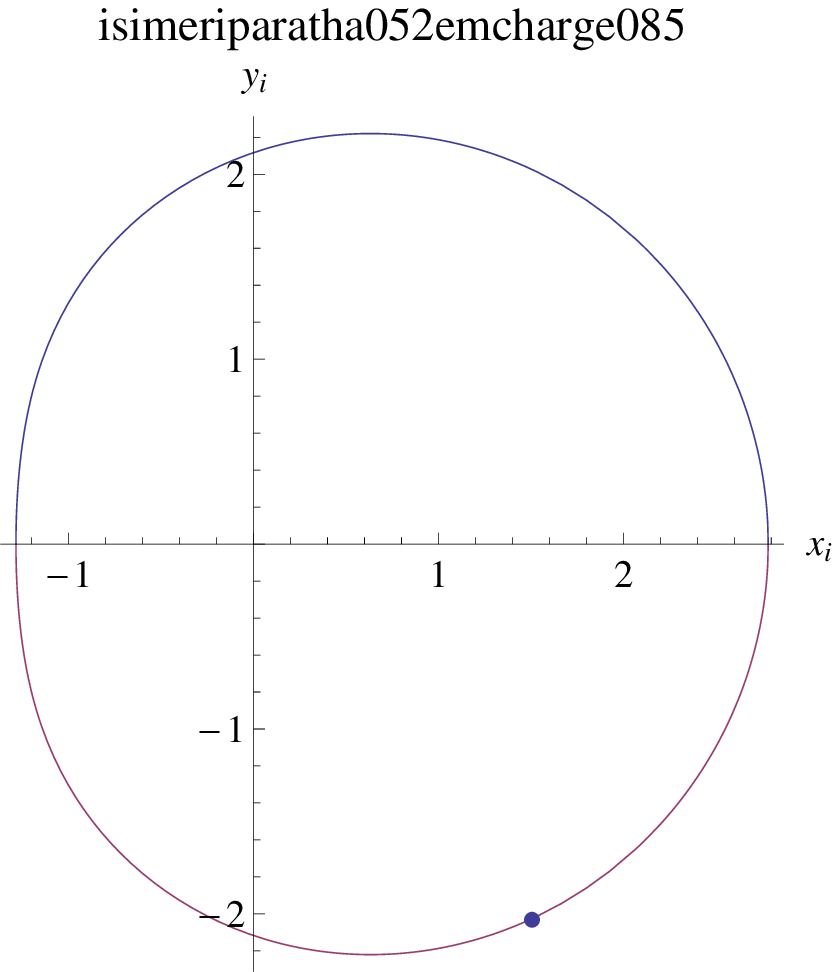}
\caption{The boundary of the shadow of the charged rotating KN black hole for Kerr parameter
 $a=0.52$, $e=0.85$ for an equatorial observer, i.e. located at polar position $\theta_O=\pi/2$. With the blue dot
 we exhibit the image solution of the KN lens equations, see Table 9.}%
\label{EMSWKN5285PI2}%
\end{center}
\end{figure}

\begin{figure}
[ptbh]
\psfrag{equatoobserva052emcharge011}{$a=0.52,e=0.11,\theta_O=\frac{\pi}{2}$}
\begin{center}
\includegraphics[height=2.3386in, width=2.1386in ]{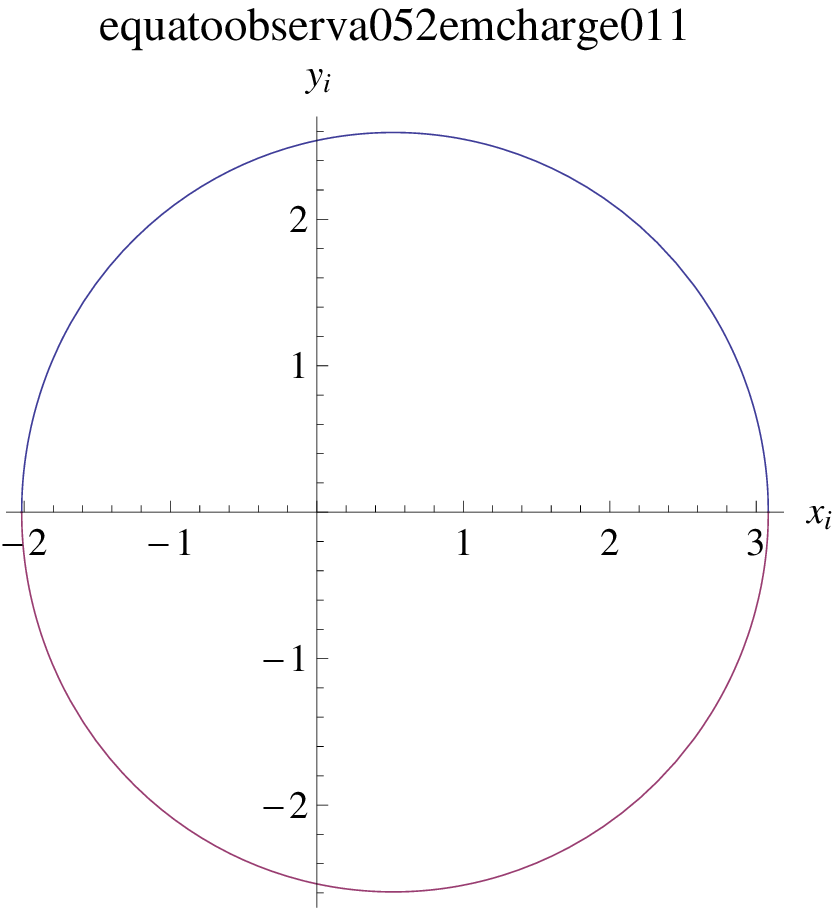}
\caption{The boundary of the shadow of the charged rotating KN black hole for Kerr parameter
 $a=0.52$, $e=0.11$ for an equatorial observer, i.e. located at polar position $\theta_O=\pi/2$.}%
\label{EMSWKN5211PI2}%
\end{center}
\end{figure}

By plugging equations $($\ref{arxikescondipcc}$)$ $\ $into
equations $(\ref{observationKNs}),$ we derive the coordinates on
the observer's image plane at which the escaped photon will be
detected for the case of a
Kerr-Newman gravitational lens:%
\begin{align}
x_{i}  &
=\frac{1}{r_{O}\sin\theta_{O}}\frac{[a^{2}(r+M)+2e^{2}r-3Mr^{2}+r^{3}]}{a\left(
r-M\right)
},\nonumber\\
y_{i}  & =\pm%
\Biggl\{%
-r^{2}\left[  4a^{2}(e^{2}-Mr)+\left(2e^{2}+r\left(
-3M+r\right)\right)^{2}\right] \nonumber\\
& -2ra^{2}z_{O}\left[ r^{3}-3rM^{2}+2a^{2}M+2e^{2}M\right] \nonumber\\
& -a^{4}z_{O}^{2}\left(  r-M\right)  ^{2}%
\Biggr\}%
^{1/2}%
\Big/%
[r_{O}^{2}\sin^{2}\theta_{O}a^{2}(r-M)^2]^{1/2}\label{iskiosfortismenisbh}%
\end{align}
For zero electric charge, i.e for $e=0,$ equations $(\ref{iskiosfortismenisbh}%
)$ reduce correctly to the coordinates on the observer's image
plane of an escaped photon from an uncharged rotating (Kerr) black
hole, eqns. ($28$) in \cite{KraniotisGVlightII}. Let us look at
some examples of how the shadow of the electromagnetic rotating
black hole is perceived by an observer at different polar
positions $\theta_O$ for different sets of values for the spin and
electric charge of the KN singularity \footnote{We note at this
point that the shadow of the KN spacetime has also been studied in
\cite{deVries}. However, the author of \cite{deVries} did not
actually solved the KN lens equations and he also considered cases
which violate (\ref{FundEcGMa}).}Our results are displayed in
Figures
\ref{SWKN993911P3},\ref{SWKN5285PI3},\ref{EMROTSWKN5204PI3},\ref{EMSWKN5211PI3},\ref{EMSWKN993911PI2},
\ref{EMSWKN5285PI2},\ref{EMSWKN5211PI2}.

From our results we observe that the larger the electric charge
(for fixed black hole spin and polar position of the observer) the
smaller the boundary of the shadow of the KN black hole, see
Figures \ref{SWKN5285PI3}-\ref{EMSWKN5211PI3} for an observer
located at $\theta_O=\pi/3,\phi_O=0$. See also Figures
\ref{EMSWKN5285PI2}-\ref{EMSWKN5211PI2} for an equatorial
observer. When we compare the case of a fast-spinning KN black
hole, Figure \ref{SWKN993911P3} with the corresponding uncharged
rotating black hole, Figure 4, page 25 of
\cite{KraniotisGVlightII}, we also observe that the charged
fast-spinning black hole has a slightly smaller boundary for its
shadow. The significant deformation of the boundary of the shadow
from circularity is present in both KN and Kerr fast spinning
black holes.

\section{Exact calculation of radial integrals for generic orbits in
Kerr-Newman spacetime}

We now perform the radial integration for generic orbits, i.e. orbits for
which both $\Phi,\mathcal{Q}$ differ from zero, assuming $\Lambda=0.$ The
relevant differential equation for the second KN lens equation that determines
the azimuth angle is:%
\begin{equation}
\frac{\mathrm{d}\phi}{\mathrm{d}r}=\frac{a(-e^{2}+2Mr)}{\pm\Delta^{KN}\sqrt{R}}
+\frac{-a^{2}\Phi}{\pm\Delta^{KN}\sqrt{R}}+\frac{\Phi}%
{\pm\sin^{2}\theta\sqrt{\Theta}}\frac{{\rm d}\theta}{{\rm d}r}.\label{lens2KN}%
\end{equation}
Thus for the radial contribution we need to integrate the first two terms.

For an observer and a source located far away from the black hole, the
relevant radial integrals can take the form:%

\begin{equation}
\fbox{$\displaystyle\int^r\rightarrow-\int_{r_S}^{\alpha}+\int_{\alpha}%
^{r_O}\simeq2\int_{\alpha}^{\infty}$}
\end{equation}
Thus we must compute the radial integral:
\begin{equation}
\Delta\phi_{rgoKN}^{GTR}=2\int_{\alpha}^{\infty}\frac{a(-e^{2}+2Mr
)-a^{2}\Phi}{\Delta^{KN}\sqrt{R}}{\rm d}r
\end{equation}

Using partial fractions we compute the previous integral in closed
analytic
form in terms of Lauricella's and Appell's generalized hypergeometric functions:%

\begin{align}
\Delta\phi_{rgoKN}^{GTR}  & =2%
\Biggl[%
\frac{-2A_{+}^{goKN}\sqrt{\omega}(\alpha_{\mu+1}-\alpha_{\mu+2})}{H^{+}}%
F_{D}\left(  \frac{1}{2},\mathbf{\beta}_{\mathbf{3}}^{\mathbf{9}},\frac{3}%
{2},\mathbf{z}_{+}^{\mathbf{r}}\right) \nonumber\\
& +\frac{A_{+}^{goKN}\sqrt{\omega}(\alpha_{\mu+1}-\alpha_{\mu+2})}{H^{+}}%
\Biggl(%
-\frac{1}{\kappa_{+}^{\prime2}}F_{1}\left(  \frac{1}{2},\mathbf{\beta}%
_{A}^{\mathbf{ra}},\frac{3}{2},\mathbf{z}_{\mathbf{A}}^{\mathbf{r}}\right)
2\nonumber\\
& +\frac{1}{\kappa_{+}^{\prime2}}F_{D}\left(  \frac{1}{2},\mathbf{\beta
}_{\mathbf{3}}^{\mathbf{9}},\frac{3}{2},\mathbf{z}_{+}^{\mathbf{r}}\right)  2%
\Biggr)%
\nonumber\\
& +\frac{-2A_{-}^{goKN}\sqrt{\omega}(\alpha_{\mu+1}-\alpha_{\mu+2})}{H^{-}%
}F_{D}\left(  \frac{1}{2},\mathbf{\beta}_{\mathbf{3}}^{\mathbf{9}},\frac{3}%
{2},\mathbf{z}_{-}^{\mathbf{r}}\right) \nonumber\\
& +\frac{A_{-}^{goKN}\sqrt{\omega}(\alpha_{\mu+1}-\alpha_{\mu+2})}{H^{-}}%
\Biggl(%
-\frac{1}{\kappa_{-}^{\prime2}}F_{1}\left(  \frac{1}{2},\mathbf{\beta}%
_{A}^{\mathbf{ra}},\frac{3}{2},\mathbf{z}_{\mathbf{A}}^{\mathbf{r}}\right)
2\nonumber\\
& +\frac{1}{\kappa_{-}^{\prime2}}F_{D}\left(  \frac{1}{2},\mathbf{\beta
}_{\mathbf{3}}^{\mathbf{9}},\frac{3}{2},\mathbf{z}_{-}^{\mathbf{r}}\right)  2%
\Biggr)%
\Biggr]%
\equiv R_{2}^{KN}(x_{i},y_{i}).\nonumber\\
\end{align}
where
\begin{equation}
A_{\pm}^{goKN}=\frac{\pm2Mar_{\pm}\mp
a(e^{2}+a\Phi)}{r_{+}-r_{-}}.
\end{equation}
The radial polynomial now has the form:%
\begin{equation}
R(r)=r^{4}+r^{2}(a^{2}-\Phi^{2}-\mathcal{Q})+2M(\mathcal{Q}%
+(\Phi-a)^{2})r-e^{2}(\mathcal{Q}+(\Phi-a)^{2})-a^{2}\mathcal{Q}.
\end{equation}
Its four roots are solved in closed analytic form by the
Weierstra\ss \ function and its derivative-see equations
(\ref{maxweierstrass})-(\ref{fourth})- where the Weierstra\ss \
invariants are given now in terms of the spin, electric charge of
the black hole, and the impact parameter and Carter's
constant, by the equations:%
\begin{align}
g_{2}  & =\frac{1}{12}(a^{2}-\Phi^{2}-\mathcal{Q)}^{2}-e^{2}(\mathcal{Q}%
+(\Phi-a)^{2})-a^{2}\mathcal{Q},\label{invg2genereicorbit}\\
g_{3}  & =-\frac{1}{216}(a^{2}-\Phi^{2}-\mathcal{Q)}^{3}-\frac{1}{4}[\mathcal{Q}+(\Phi-a)^{2}]^{2}\nonumber\\
& +[-e^{2}(\mathcal{Q}+(\Phi-a)^{2})-a^2\mathcal{Q}]\left(  \frac{a^{2}-\Phi^{2}-\mathcal{Q}%
}{6}\right) \label{genericKarlWr}
\end{align}
while
\begin{equation}
x_{1}=\wp^{-1}\left(  -\frac{a^{2}-\Phi^{2}-\mathcal{Q}}{6}\right)
\label{genpointx1} .
\end{equation}

We have written the lens equations $(\ref{lensKN1})$ and
$(\ref{lens2KN})$ in
the form:%
\begin{align}
R_{1}^{KN}(x_{i},y_{i})-A_{1}(x_{i},y_{i},x_{S},y_{S},m)  & =0,\\
\Delta\phi(x_{S},y_{S},n)-R_{2}^{KN}(x_{i},y_{i})-A_{2}(x_{i},y_{i}%
,x_{S},y_{S},m)  & =0.
\end{align}
In these equations $n$ denotes the number of windings around the
$z$-axis and $m$ the number of turning points in the polar motion.
The term $R_{1}^{KN}(x_{i},y_{i})$ in closed analytic form is
given by equation:
\begin{equation}
2\int_{\alpha}^{\infty}\frac{\mathrm{d}r}{\sqrt{R}}=2\frac{\Gamma
(1/2)\Gamma(1)}{\Gamma(3/2)}\left( \frac{1}{\sqrt{(\gamma-\alpha
)(\delta-\alpha)}}\right)  F_{1}\left(  \frac{1}{2},\mathbf{\beta}%
_{A}^{\mathbf{ra}},\frac{3}{2},\mathbf{z}_{\mathbf{A}}^{\mathbf{r}}\right)
\equiv R_{1}^{KN}(x_{i},y_{i})
\end{equation}
and equations (\ref{maxweierstrass})-(\ref{fourth}),
(\ref{invg2genereicorbit})-(\ref{genpointx1}). The angular parts
of the
lens equations: $A_{1}(x_{i},y_{i},x_{S},y_{S},m),$ $A_{2}(x_{i},y_{i}%
,x_{S},y_{S},m)$ are the same as in the uncharged case (the Kerr field) and
have been computed analytically in \cite{KraniotisGVlightII} for the case of a
light trajectory that encounters $m$ turning points $(m\geq1)$ in the polar
motion:

\begin{equation}
\fbox{$ \displaystyle\pm\int_{\theta_S}^{\theta_{\rm{min/max}}}
\underbrace{\pm\int_{\theta_{\rm{min/max}}}^{\theta_{\rm{max/min}}}
\pm\int_{\theta_{\rm{max/min}}}^{\theta_{\rm{min/max}}}\cdots}_{m-1\;{\rm
times}}\pm\int_{\theta_{\rm{max/min}}}^{\theta_O}$}
\end{equation}%
 Combining the exact results for the radial integrals we have computed in
this work with the angular integral computations of \cite{KraniotisGVlightII}
for the Kerr black hole, which remain valid, as we said in the KN case we have
that our exact solutions of the lens equations for a KN black hole are:%

\begin{eqnarray}
R_1^{KN}(x_i,y_i)
&=&A_{1}(x_{i},y_{i},x_{S},y_{S},m)\Leftrightarrow\frac{2}{\sqrt
{(\alpha-\gamma)(\alpha-\delta)}}\frac{\Gamma(1/2)}{\Gamma(3/2)}F_{1}\left(
\frac{1}{2},\mbox{\boldmath${\beta}^{ra}_A$},\frac{3}{2},{\bf{z_{A}^{r}}%
}\right)   \notag\\
&=&2(m-1)\frac{1}{2|a|}\sqrt{\frac{z_{m}}{z_{m}(z_{m}-z_{3})}}\pi F\left(
\frac{1}{2},\frac{1}{2},1,\frac{z_{m}}{z_{m}-z_{3}}\right)   \notag\\
&+&\frac{1}{2|a|}\frac{\sqrt{(z_{m}-z_{S})}}{\sqrt{z_{m}(z_{m}-z_{3})}%
}F_{1}\left( \frac{1}{2},\mbox{\boldmath${\beta}^{ra}_A$},\frac{3}{2}%
,{\bf{z_A^{1S}}}\right) \frac{\Gamma(\frac{1}{2})\Gamma(1)}{\Gamma(3/2)}
\notag\\
&+&[1-\mathrm{sign(}\theta_{S}\circ\theta_{mS})]\frac{1}{|a|}\frac
{\sqrt{\frac{z_{S}(z_{m}-z_{3})}{z_{m}(z_{S}-z_{3})}}}{\sqrt{z_{m}%
-z_{3}}}\times F_{1}\left(
\frac{1}{2},\mbox{\boldmath${\beta}^{ra}_A$},\frac{3}{2},{\bf{z_A^{2S}}}%
\right)   \notag\\
&+&\frac{1}{2|a|}\frac{\sqrt{(z_{m}-z_{O})}}{\sqrt{z_{m}(z_{m}-z_{3})}%
}F_{1}\left( \frac{1}{2},\mbox{\boldmath${\beta}^{ra}_A$},\frac{3}{2}%
,{\bf{z_A^{1O}}}\right) \frac{\Gamma(\frac{1}{2})\Gamma(1)}{\Gamma(3/2)}
\notag\\
&+&[1-\mathrm{sign(}\theta_{O}\circ\theta_{mO})]\frac{1}{|a|}\frac
{\sqrt{\frac{z_{O}(z_{m}-z_{3})}{z_{m}(z_{O}-z_{3})}}}{\sqrt{z_{m}%
-z_{3}}}\times F_{1}\left(
\frac{1}{2},\mbox{\boldmath${\beta}^{ra}_A$},\frac{3}{2},{\bf{z_A^{2O}}}%
\right), \notag\\
\label{Constraint1one}
\end{eqnarray}%
\begin{equation}
-\phi_{S}=R_{2}^{KN}(x_{i},y_{i})+A_{2}(x_{i},y_{i},x_{S},y_{S}%
,m),\label{fakos2}%
\end{equation}%
\begin{equation}
\xi_{S}=\wp\left(2\sigma_S\left[  R_{1}^{KN}(x_{i},y_{i})-2(m-1)\frac{1}{2|a|}\sqrt%
{\frac{z_{m}}{z_{m}(z_{m}-z_{3})}}\pi F\left(  \frac{1}{2},\frac{1}{2}%
,1,\frac{z_{m}}{z_{m}-z_{3}}\right)  +\cdots\right]+\epsilon\right)
\label{combinlens}%
\end{equation}
where%
\begin{eqnarray}
A_{2}(x_{i},y_{i},x_{S,}y_{S},m) &=&2(m-1)\times\left[ \frac{\Phi}{|a|}%
\frac{1}{\sqrt{(z_{m}-z_{3})}}\frac{1}{(1-z_{m})}\frac{\pi}{2}F_{1}%
\left( \frac{1}{2},1,\frac{1}{2},1,\frac{-z_{m}}{1-z_{m}},\frac{z_{m}}%
{z_{m}-z_{3}}\right) \right]  \notag\\
&+&\frac{\Phi}{2|a|}\sqrt{\frac{(z_{m}-z_{S})}{z_{m}}}\frac{1}%
{\sqrt{z_{m}-z_{3}}}\frac{2}{(1-z_{m})}\times F_{D}\left(
\frac{1}{2},\mbox{\boldmath${\beta}^4_3$},\frac{3}{2},{\bf{z_S^1}}\right)
\notag\\
&+&[1-\mathrm{sign(}\theta_{S}\circ\theta_{mS}\mathrm{)]}\frac{\Phi}{|a|}%
\frac{z_{S}}{z_{m}}\frac{z_{S}-z_{m}}{1-z_{S}}\frac{1}{\sqrt{z_{S}%
(z_{S}-z_{m})(z_{3}-z_{S})}}\times\notag\\
&&F_{D}\left( 1,\mbox{\boldmath${\beta}^7_3$},\frac{3}{2}, {\bf{z_S^2}}\right)
+  \notag\\
&+&\frac{\Phi}{2|a|}\sqrt{\frac{(z_{m}-z_{O})}{z_{m}}}\frac{1}%
{\sqrt{z_{m}-z_{3}}}\frac{2}{(1-z_{m})}\times F_{D}\left(
\frac{1}{2},\mbox{\boldmath${\beta}^4_3$},\frac{3}{2},{\bf{z_O^1}}\right)
\notag\\
&+&[1-\mathrm{sign(}\theta_{O}\circ\theta_{mO}\mathrm{)]}\frac{\Phi}{|a|}%
\frac{z_{O}}{z_{m}}\frac{z_{O}-z_{m}}{1-z_{O}}\frac{1}{\sqrt{z_{O}%
(z_{O}-z_{m})(z_{3}-z_{O})}}\times\notag\\
&&F_{D}\left( 1,\mbox{\boldmath${\beta}^7_3$},\frac{3}{2},{\bf{z_O^2}}\right)
\notag\\
&&  \label{KleistiGwniaki2}
\end{eqnarray}%
and:
\begin{equation}
\fbox{$\displaystyle \theta_{mO}:={\rm Arccos}({\rm
sign}(y_i)\sqrt{z_m})= {\rm Arccos}({\rm
sign}(\beta_i)\sqrt{z_m}), $}
\end{equation}
$y_{i}$ is the possible position of the image and:
\begin{equation}
\theta _{mS}:=\left\{
\begin{array}{lll}
\theta _{mO}, & m & \mathrm{odd} \\
\pi -\theta _{mO}, & m & \mathrm{even}%
\end{array}%
\right.
\end{equation}
Also $\theta_1\circ\theta_2:=\cos\theta_1\cos\theta_2$,
$\sigma_S:=\mathrm{sign} \;\theta_S\circ\theta_{mS}$ and
$\epsilon$ denotes a constant of integration. The Weierstra\ss \
invariants in equation $(\ref{combinlens})$ are defined in
$(\ref{CarlWeierstrassInstitut})$ and $(\ref{karlhumboldtinv})$
\footnote{In establishing (\ref{combinlens}) we used the fact that
the sum of the second and third term on the right hand side of
eq.(\ref{Constraint1one}) can be written as:
$\int_{\xi_S}^{\xi_m}+(1-\sigma_S)\int_{\xi_0}^{\xi_S}$, where
$(\xi_m,\xi_0)$ are extremal values of $\xi$; thus, one can
separate from it the expression
$\sigma_S\int_{\xi_S}^{\infty}\propto\sigma_S \wp^{-1}(\xi)$.}.

A solution of the KN lens equations with $m=3$ is presented in
Table \ref{LENSKN1s} for $a=0.52,e=0.85$. The solution as it
appears on the observer's image plane is exhibited along with the
boundary of the shadow of the KN black hole in
Fig.\ref{EMSWKN5285PI2}. Another solution of the KN lens equations
for an electromagnetic fast spinning black hole
($a=0.9939,e=0.11$), is presented in table \ref{LENSKN2s}. The
solution as it appears on the observer's image plane is exhibited
along with the boundary of the shadow of the KN black hole in
Fig.\ref{EMSWKN993911PI2}.

\begin{table}[tbp] \centering
\begin{tabular}
[c]{|l|l|}\hline
\textbf{Solution with Parameters:} & $a=0.52,e=0.85,Q=16.51343,\Phi=-3.0120$%
\\\hline
$ \alpha_i\left(\frac{GM}{c^2}\right)$, & $3.0120$\\
 $\beta_i\left(\frac{GM}{c^2}\right)$ & $-4.06367$ \\
 $x_i\left(\frac{2}{r_O}\frac{GM}{c^2}\right)$& $1.506$\\
 $y_i\left(\frac{2}{r_O}\frac{GM}{c^2}\right)$& $-2.031835$ \\
 $m$&$3$ \\
 $z_S$ &$0.06859659416$ \\
 $\theta_S$ & $74.82^{\circ}$ \\
 $\Delta\phi({\rm rad})$ & $-8.39801$ \\
 $\phi_S$ & $121.17^{\circ}$ \\
 $\omega$ & $0.6211809022$ \\
 $\omega^{\prime}$ & $1.5336366498i$
\\\hline
\end{tabular}
\caption{Solution of the lens equations in the Kerr-Newman geometry and the predictions for the source and image positions
for an observer at $\theta_O=\pi/2, \phi_O=0$. The number of turning points in the polar variable is 3. The values for the
Kerr parameter, impact factor are in units of $\frac{GM}{c^2}$, those of electric charge in units of $\sqrt{G}M_{BH}$ and of
Carter's constant in units of $\frac{G^2M^2}{c^4}$.}\label{LENSKN1s}%
\end{table}%

\begin{table}[tbp] \centering
\begin{tabular}
[c]{|l|l|}\hline
\textbf{Solution with Parameters:} & $a=0.9939,e=0.11,Q=25.790421,\Phi=-3.0118023$%
\\\hline
$ \alpha_i\left(\frac{GM}{c^2}\right)$, & $3.0118023$\\
 $\beta_i\left(\frac{GM}{c^2}\right)$ & $-5.07843$ \\
 $x_i\left(\frac{2}{r_O}\frac{GM}{c^2}\right)$& $1.5059$\\
 $y_i\left(\frac{2}{r_O}\frac{GM}{c^2}\right)$& $-2.53921$ \\
 $m$&$3$ \\
 $z_S$ &$0.3110425032$ \\
 $\theta_S$ & $56.10^{\circ}$ \\
 $\Delta\phi({\rm rad})$ & $-7.92181$ \\
 $\phi_S$ & $93.8862^{\circ}$ \\
 $\omega$ & $0.5312062705$ \\
 $\omega^{\prime}$ & $1.1216689490i$
\\\hline
\end{tabular}
\caption{Solution of the lens equations in the Kerr-Newman
geometry and the predictions for the source and image positions
for an observer at $\theta_O=\pi/2, \phi_O=0$. The number of
turning points in the polar variable is 3. The values for the Kerr
parameter, impact factor are in units of $\frac{GM}{c^2}$, those
of electric charge in units of $\sqrt{G}M_{BH}$ and of
Carter's constant in units of $\frac{G^2M^2}{c^4}$.}\label{LENSKN2s}%
\end{table}%

A detailed analysis of gravitational lensing in the KN and KN(a)dS
spacetimes will be a subject of a separate publication. We also
leave for the future the exact analytic computation of the
magnification factors for the KN and KN(a)dS spacetimes.

\section{Conclusions\label{symperasma}}

We have solved in closed analytic form the null geodesic equations
in Kerr-Newman and Kerr-Newman-(anti) de Sitter spacetimes. The
analytic solutions were expressed elegantly in terms of
generalized hypergeometric functions of Lauricella and Appell as
well as the Weierstra\ss \ elliptic function.

We also solved the more involved problem of treating a Kerr-Newman
black hole as a gravitational lens, i.e. a KN black hole along
with a static source of light and a static observer both located
far away but otherwise at arbitrary positions in space. Again, for
this model we give the analytic solutions of the lens equation in
terms of Appell and Lauricella hypergeometric functions and the
Weierstra$\ss$  modular form.

 We applied our exact solutions for
the calculation of the frame dragging effect for spherical polar
and non-polar photon orbits in the gravitational field of a KN
black hole. We also applied our exact solution for the
gravitational bending of light that an equatorial unbound photon
orbit experiences in the curved spacetime of a charged rotating
black hole. We noted the significant dependence of the deflection
angle on the electric charge of the spacetime singularity in
regions of the parameter space. This result, in conjuction with
our solution for the periapsis advance of a neutral test particle
in an equatorial non-circular orbit in KN spacetime and its
application to the observed orbits of $S$-stars, indicates that
future measurements of the galactic centre black hole and its
relativistic observables may constrain significantly or detect the
electric charge of the galactic rotating black hole.

We also derived analytic expressions for the Maxwell tensor
components for a ZAMO frame in the KNdS spacetime.

Future directions of research will include the application of the
closed form analytic solution of the lens equations in the
Kerr-Newman family of spacetimes to the important case of the
SgrA* supermassive black hole.

Another interesting avenue of research will be the application of
our exact solutions in the exciting field of $e^{-}-e^{+}$ pair
creation by vacuum polarization around electromagnetic black holes
and the theory of pulsars. Indeed, the Kerr-Newman electromagnetic
field is such that $^{\ast}F^{\mu\nu }F_{\mu\nu}\neq0$ or in
pulsar language $\mathbf{E\cdot B\neq}0$ \cite{PUNSLY}$.$ It seems
that the KN black hole is charged just like the neutron star in
pulsar models \cite{Ruderman}.

This angle of research will have potentially very important
applications for the gamma ray bursts from electromagnetic black
holes \cite{Cherubini} and pulsars as well as in the study of the
vacuum structure of non-linear electrodynamics \cite{QEDVac}. The
study of the Faraday effect in the KN-(a)dS electromagnetic black
hole will serve nicely towards such investigations. We aim to
pursue this exciting interplay of relativistic astrophysics and
non-linear electrodynamics in the future
\cite{KRANIOTISGVQEDPEMBH}.

Thus the theory we developed in this work based on the exact
solutions of null geodesics in the KNdS spacetime will help in
exploring the strong field regime of General Relativity and the
structure of Electrodynamics in curved spacetimes.

\begin{acknowledgement}
I thank my family  for support, Zden\u{e}k Stuchl\'{\i}k,
 G.
Leontaris and M. Stamatis for discussions. I am grateful to the referees for the careful reading of the manuscript and their very constructive comments and suggestions that helped improve the presentation of this work.
\end{acknowledgement}

\appendix{}

\section{The roots of the quartic in terms of Weierstra\ss \ functions}

We are going to use the addition theorem for the Weierstra\ss \ elliptic
function to express the roots of the quartic $P_{4}(x)=x^{4}+ax^{2}+bx+c\in%
\mathbb{C}
\lbrack x]$ in terms of the Weierstra\ss \ functions following \cite{McKean}.
We first write $x$ for a point of the cover $\mathbb{C}
$ and $\mathfrak{p}=({\bf x}\mathbf{,{\bf y}})$ for the corresponding point of the cubic determined by ${\bf x}=\wp(x)$ and $\mathbf{{\bf y}%
=\wp}^{\prime}(x).$ Then we study the intersections of the cubic $\mathbf{%
{\bf y}}^{2}=4{\bf x}^{3}-g_{2}{\bf x}-g_{3}$ and the line $\mathbf{{\bf y}}=a{\bf x}+b:$ ${\bf x}_{1},{\bf x}_{2},{\bf x}_{3}$ are the roots of
\begin{align}
F({\bf x})  & =4
{\bf x}^{3}-g_{2}{\bf x}-g_{3}-(a{\bf x}+b)^{2}\nonumber\\
& =4(
{\bf x}-{\bf x}_{1})({\bf x}-{\bf x}_{2})({\bf x}-{\bf x}_{3})
\end{align}
so
\begin{align}
4({\bf x}_{1}-{\bf x}_{2})({\bf x}_{1}-{\bf x}_{3})  & =F^{\prime}({\bf x}_{1})=12{\bf x}_{1}^{2}-g_{2}-2a(a{\bf x}_{1}+b)\nonumber\\
& =12{\bf x}_{1}^{2}-g_{2}-2a\mathbf{{\bf y}}_{1}\mathbf{.\label{firsteqnar}}%
\end{align}
and
\begin{equation}
({\bf x}_{2}-{\bf x}_{1})+({\bf x}_{3}-{\bf x}_{1})={\bf x}_{1}+{\bf x}_{2}+{\bf x}_{3}-3{\bf x}_{1}=\frac{a^{2}}{4}-3{\bf x}_{1}.\label{secondarteqn}%
\end{equation}
We note that $a$ is the slope of the line, so for distinct ${\bf x}_{1},{\bf x}_{2},{\bf x}_{3}$ we have
\begin{equation}
a=\frac{\mathbf{{\bf y}}_{2}-\mathbf{{\bf y}}_{1}}{{\bf x}_{2}-{\bf x}_{1}}=\frac{\mathbf{{\bf y}}_{3}-\mathbf{{\bf y}}_{2}}{{\bf x}_{3}-{\bf x}_{2}}=\frac{\mathbf{{\bf y}}_{1}-\mathbf{{\bf y}}_{3}}{{\bf x}_{1}-{\bf x}_{3}}.
\end{equation}
Now%
\begin{align}
& (\ref{secondarteqn})^{2}-(\ref{firsteqnar})\nonumber\\
& =({\bf x}_{2}+{\bf x}_{3}-2{\bf x}_{1})^{2}-4{\bf x}_{1}^{2}+4{\bf x}_{1}{\bf x}_{3}+4{\bf x}_{1}{\bf x}_{2}-4{\bf x}_{2}{\bf x}_{3}\nonumber\\
& =({\bf x}_{2}-{\bf x}_{3})^{2}=\left(  \frac{a^{2}}{4}-3{\bf x}_{1}\right)  ^{2}-12
{\bf x}_{1}^{2}+g_{2}+2a\mathbf{{\bf y}}_{1}\nonumber\\
& =\left(  \frac{a}{2}\right)  ^{4}-6{\bf x}_{1}\left(  \frac{a}{2}\right)  ^{2}+4\mathbf{{\bf y}}_{1}\left(  \frac{a}{2}\right)  -3{\bf x}_{1}^{2}+g_{2}.\\
& \nonumber
\end{align}
Thus%
\begin{align}
X  & =\frac{a}{2}=\frac{1}{2}\frac{\mathbf{{\bf y}}_{2}-\mathbf{{\bf y}}_{1}}{{\bf x}_{2}-{\bf x}_{1}}=\frac{1}{2}\frac{\wp^{\prime}(x_{2})-\wp^{\prime}(x_{1})}{\wp(x_{2}%
)-\wp(x_{1})},\\
Y  & ={\bf x}_{3}-{\bf x}_{2}=\wp(-x_1-x_2)-\wp(x_2)=\wp(x_{1}+x_{2})-\wp(x_{2})\Rightarrow
\label{ARTIAWP}
\end{align}%
\begin{equation}
Y^{2}=X^{4}-6\wp(x_{1})X^{2}+4\wp^{\prime}(x_{1})X-3\wp^{2}(x_{1})+g_{2}\equiv
P_{4}(X).
\end{equation}
In the second equality of (\ref{ARTIAWP}) we used the addition theorem $x_1+x_2=-x_3$ in $\mathbb{C}/\mathbb{L}$, $\mathbb{L}$ the period lattice, and the fact that the Weierstra$\ss$ function is even.
For fixed $x_{1},$ and variable $x_{2},$ $P_{4}(X)=0$ only if $Y=0.$ This
occurs for $x_{2}=-\frac{x_{1}}{2}$ to which may be added one of the three
half-periods producing four roots of $P_{4}(x)=0$ and these must be distinct, see equations (\ref{maxweierstrass})-(\ref{fourth}).

\section{Lauricella's multivariable hypergeometric function $F_D$}
In this appendix B, we define  Lauricella's $4^{th}$
hypergeometric function of $m$-variables and its integral
representation:

\begin{equation}
\fbox{$\displaystyle
F_D(\alpha,\mbox{\boldmath${\beta}$},\gamma,{\bf z})=
\sum_{n_1,n_2,\dots,n_m=0}^{\infty}\frac{(\alpha)_{n_1+\cdots
n_m}(\beta_1)_{n_1} \cdots (\beta_m)_{n_m}}
{(\gamma)_{n_1+\cdots+n_m}(1)_{n_1}\cdots (1)_{n_m}}
z_1^{n_1}\cdots z_m^{n_m}$} \label{GLauri}
\end{equation}%

where
\begin{eqnarray}
\mathbf{z} &=&(z_{1},\ldots ,z_{m}),  \notag \\
\mbox{\boldmath${\beta}$} &=&(\beta _{1},\ldots ,\beta _{m}).
\label{TUPLES}
\end{eqnarray}

The Pochhammer symbol
\fbox{$\displaystyle (\alpha)_m=(\alpha,m)$}
is defined by%
\begin{equation}
(\alpha )_{m}=\frac{\Gamma (\alpha +m)}{\Gamma (\alpha )}=\left\{
\begin{array}{ccc}
1, &
{\rm if}
& m=0 \\
\alpha (\alpha +1)\cdots (\alpha +m-1) & \text{%
{\rm if}%
} & m=1,2,3%
\end{array}%
\right.
\end{equation}
With the notations $\mathbf{z^n}:=z_1^{n_1}\cdots z_m^{n_m}$,
$(\mbox{\boldmath${\beta}$})_{\mathbf{n}}:=(\beta_1)_{n_1}\cdots
(\beta_m)_{n_m}$, $\mathbf{n!}=n_1!\cdots n_m!$,
$|\mathbf{n}|:=n_1+\cdots+n_m$ for $m$-tuples of numbers in
(\ref{TUPLES}) and of non-negative integers
$\mathbf{n}=(n_1,\cdots n_m)$ the Lauricella series $F_D$ in
compact form is
\begin{equation}
F_D(\alpha,\mbox{\boldmath${\beta}$},\gamma,{\bf
z}):=\sum_{\mathbf{n}}\frac{
(\alpha)_{|\mathbf{n}|}(\mbox{\boldmath${\beta}$})_{\mathbf{n}}}{(\gamma)_{|\mathbf{n}|}\mathbf{n!}}\mathbf{z^n}
\end{equation}

The series admits the following integral representation:

\begin{equation}
\fbox{$\displaystyle
F_D(\alpha,\mbox{\boldmath${\beta}$},\gamma,{\bf z})=
\frac{\Gamma(\gamma)}{\Gamma(\alpha)\Gamma(\gamma-\alpha)}
\int_0^1 t^{\alpha-1}(1-t)^{\gamma-\alpha-1}(1-z_1
t)^{-\beta_1}\cdots (1-z_m t)^{-\beta_m} {\rm d}t $}
\label{OloklAnapa}
\end{equation}
which is valid for
\fbox{$\displaystyle {\rm Re}(\alpha)>0,\;{\rm Re}(\gamma-\alpha)>0. $}%
. It
{\em converges\;absolutely}
inside the m-dimensional cuboid:%
\begin{equation}
|z_{j}|<1,(j=1,\ldots ,m).
\end{equation}
For $m=2$ $F_D$ in the notation of Appell becomes the two variable
hypergeometric function
$F_1(\alpha,\beta,\beta^{\prime},\gamma,x,y)$ with integral
representation:
\begin{equation}
\int_0^1 u^{\alpha-1}(1-u)^{\gamma-\alpha-1}(1-u x)^{-\beta}(1-u
y)^{-\beta^{\prime}}{\rm
d}u=\frac{\Gamma(\alpha)\Gamma(\gamma-\alpha)}{\Gamma(\gamma)}
F_1(\alpha,\beta,\beta^{\prime},\gamma,x,y)
\end{equation}

\section{Calculation of the Maxwell tensor components in the ZAMO frame for the KNdS spacetime}
The ZAMO basis vectors are determined by the transformation \
\begin{equation}
\hat{e}_0=|g_{tt}-\Omega^2
g_{\phi\phi}|^{-1/2}\frac{\partial}{\partial
t}+\Omega|g_{tt}-\Omega^2
g_{\phi\phi}|^{-1/2}\frac{\partial}{\partial \phi},
\end{equation}

\begin{equation}
\hat{e}_{\phi}=\frac{1}{\sqrt{g_{\phi\phi}}}\frac{\partial}{\partial
\phi},
\end{equation}
\begin{equation}
\hat{e}_r=\left(\frac{\Delta_r^{KN}}{\rho^2}\right)^{1/2}\frac{\partial}{\partial
r},
\end{equation}
\begin{equation}
\hat{e}_{\theta}=\left(\frac{\Delta_{\theta}}{\rho^2}\right)^{1/2}\frac{\partial}{\partial
\theta},
\end{equation}
where the angular velocity $\Omega$ is given in the KNdS case by
the expression:
\begin{equation}
\Omega=\frac{-ac
\sin^2\theta[\Delta^{KN}_r-\Delta_{\theta}(r^2+a^2)]}{\Xi^2 \rho^2
g_{\phi\phi}}=\frac{-ac
[\Delta^{KN}_r-\Delta_{\theta}(r^2+a^2)]}{(\Delta_{\theta}(r^2+a^2)^2-a^2\sin^2\theta\Delta_{r}^{KN})}
\end{equation}
and the quantity $g_{tt}g_{\phi\phi}-g^2_{\phi t}=\frac{-c^2
\Delta_r^{KN}\Delta_{\theta}\sin^2\theta}{\Xi^4}$.
In addition, we compute for the lapse function $\alpha_{ZAMO}$:
\begin{equation}
\fbox{$\displaystyle \alpha_{ZAMO}:=|g_{tt}-\Omega^2
g_{\phi\phi}|^{1/2}=\frac{c(\Delta_r^{KN})^{1/2}\Delta_{\theta}^{1/2}\sin\theta}{\Xi^2
\sqrt{g_{\phi\phi}}}\label{lapsef} $} \end{equation} Equation
(\ref{lapsef}), reduces correctly, assuming $\Lambda=0$, to the lapse function derived in \cite{Thorne}.

 Now our analytic calculation for  the electric ($\vec{E}$) and magnetic fields ($\vec{B}$) in the presence
 of the cosmological constant $\Lambda$ in the ZAMO frame yields:
\begin{equation}
\shadowbox{$ \displaystyle E^{r}=\frac{(r^2+a^2)e[-r^2+a^2
\cos^2\theta]\Delta_{\theta}^{1/2}}{\sqrt{(r^2+a^2)^2
\Delta_{\theta}-a^2\sin^2\theta\Delta_r^{KN}}\rho^4},
\label{aktiElectricr} $}
\end{equation}
\begin{equation}
\shadowbox{$ \displaystyle E^{\theta}=\frac{-2era^2
(\Delta_r^{KN})^{1/2}\cos\theta\sin\theta}{\rho^4
\sqrt{(r^2+a^2)^2 \Delta_{\theta}-a^2\sin^2\theta\Delta_r^{KN}}},
\label{electheta}$}
\end{equation}

\begin{equation}
\shadowbox{$ \displaystyle  B^r=\frac{2era (r^2+a^2)\cos\theta
\Delta_{\theta}^{1/2}}{\rho^4 \sqrt{(r^2+a^2)^2
\Delta_{\theta}-a^2\sin^2\theta\Delta_r^{KN}}}, \label{magner} $}
\end{equation}

\begin{equation}
\shadowbox{$ \displaystyle
B^{\theta}=\frac{-(\Delta_r^{KN})^{1/2}a\sin\theta
e[-r^2+a^2\cos^2\theta]}{\rho^4 \sqrt{(r^2+a^2)^2
\Delta_{\theta}-a^2\sin^2\theta\Delta_r^{KN}}}.
\label{magnetictheta} $}
\end{equation}
Also it holds:
\begin{equation}
E^{\phi}=B^{\phi}=0
\end{equation}

 To the best of our knowledge equations
(\ref{aktiElectricr})-(\ref{magnetictheta}) represent the first
calculation of the Maxwell tensor components in the ZAMO frame for
the case of the Kerr-Newman-de Sitter spacetime. Our solutions for
the electric and magnetic fields, for zero cosmological constant,
reduce correctly to the corresponding expressions for the
Kerr-Newman spacetime and LNRF observers derived in \cite{Bicak}.

\end{document}